\documentclass[11pt]{article}
\usepackage{amsmath,amssymb,color,graphics,epsfig,bm}
\usepackage{graphicx}
\usepackage{subfigure}
\usepackage{amsfonts}
\newcommand{\be}{\begin{equation}}
\newcommand{\ee}{\end{equation}}
\newcommand{\bea}{\setlength\arraycolsep{2pt} \begin{eqnarray}}
\newcommand{\eea}{\end{eqnarray}}
\newcommand{\nn}{\nonumber}

\def\ft#1#2{{\textstyle{\frac{\scriptstyle #1}{\scriptstyle #2} } }}
\def\fft#1#2{{\frac{#1}{#2}}}

\def\0{{\sst{(0)}}}
\def\1{{\sst{(1)}}}
\def\2{{\sst{(2)}}}
\def\3{{\sst{(3)}}}
\def\4{{\sst{(4)}}}
\def\5{{\sst{(5)}}}
\def\6{{\sst{(6)}}}
\def\7{{\sst{(7)}}}
\def\8{{\sst{(8)}}}
\def\sst#1{{\scriptscriptstyle #1}}

\begin{document}

\begin{flushright}
%\hfill{KIAS-P12028}
 %\hfill{
%\bf hep-th/yymmnnn}
\end{flushright}

\vspace{25pt}
\begin{center}
{\large {\bf Time dependence of complexity for Lovelock black holes}}

\vspace{10pt}
 Zhong-Ying Fan$^1$ and Hua-Zhi Liang $^1$
 %Minyong Guo$^{2,3}$\\

\vspace{10pt}
$^1${ Center for Astrophysics, School of Physics and Electronic Engineering, \\
 Guangzhou University, Guangzhou 510006, China }\\
% $^2${ Department of Physics, Beijing Normal University, \\
% Beijing 100875,  P. R. China}\\
%$^3${ Perimeter Institute for Theoretical Physics \\Waterloo, Ontario N2L 2Y5, Canada\\}
\smallskip
%{\it $^{2}$Department of Physics and State Key Laboratory of Nuclear Physics and Technology,\\}
%{\it Peking University, No.5 Yiheyuan Rd, Beijing 100871, P.R. China\\}
%\smallskip
%{\it $^{3}$Collaborative Innovation Center of Quantum Matter, No.5 Yiheyuan Rd,\\}
%{\it  Beijing 100871, P. R. China\\}

\vspace{40pt}

\underline{ABSTRACT}
\end{center}
We study the general time dependence of complexity for holographic states dual to Lovelock black holes using the ``Complexity=Action" (CA) proposal. We observe that at early times, the critical time at which the complexity begins to increase is a decreasing function of the higher order coupling constants, which implies that the complexity evolves faster than that of Schwarzschild black holes. At late times, the rate of change of complexity is essentially determined by the generalised Gibbons-Hawking-York boundary term evaluated at the future singularity. In particular, its ratio to black hole mass is a characteristic constant, independent of the higher order couplings. Thus, in the vanishing coupling limit, the result in general does not reduce to that of Schwarzschild black holes, in spite of that the metric reduces to the latter as well as the gravitational action. In fact, the two differ by a constant during the whole time evolution. Including the next-to-leading order term around late times, we find that as the Einstein case, the late time limit is always approached from above, thus violating any conjectured upper bound given by the late time result. For charged Lovelock black holes, we find that with sufficient charge, the complexity roughly behaves the same as the Einstein case. However, for smaller charges, the two have some significant differences.
In particular, unlike the Einstein case, in the uncharged limit the complexity growth rate does not match with the neutral case, differing by a constant in the whole time evolution.

\vfill {\footnotesize  Email: fanzhy@gzhu.edu.cn\,.}

\thispagestyle{empty}

\pagebreak

\tableofcontents
\addtocontents{toc}{\protect\setcounter{tocdepth}{2}}

%%%%%%%%%%%%%%%%%%%%%%%%%%%%%%%%%%%%%%%%

%\newpage
%%%%%%%%%%%%%%%%%%%%%%%%%%%%%%%%%%%%%%%%

%\vspace{2cm}

\section{Introduction}
 In the past decade, entanglement entropy has become one of the most exciting topics in the AdS/CFT correspondence \cite{Maldacena:1997re,Witten:1998qj,Gubser:1998bc}. The holographic studies of entanglement entropy \cite{Ryu:2006bv} not only deepens our understanding of quantum entanglement in the boundary theories but also opens a new window towards the nature of quantum gravity \cite{VanRaamsdonk:2010pw}. However, it was argued in \cite{Susskind:2014moa} that the entanglement entropy may not be enough to probe the degrees of freedoms in black holes interior since the volume of black holes continues growing even if spacetimes reach thermal equilibrium. Instead, {\it complexity} was proposed to be the correct quantity to characterize the interior of black holes.

Complexity is an important notion in computational science. It is defined by the minimum number of gates needed to prepare a state of interest from a reference state. Roughly speaking, complexity measures how hard it is to construct a target state from a given initial (usually unentangled) state. In other words, it is a kind of distance between states in the Hilbert space. In the AdS/CFT correspondence, there are two popular proposals for complexity, usually dubbed by ``Complexity=Volume" (CV) duality \cite{Susskind:2014rva,Stanford:2014jda} and ``Complexity=Action" (CA) duality \cite{Brown:2015bva,Brown:2015lvg}. The two conjectures have been extensively studied in literature \cite{Moosa:2017yiz,HosseiniMansoori:2017tsm,Mahapatra:2018gig,Chapman:2016hwi,Carmi:2016wjl,Kim:2017lrw,Yang:2016awy,Chapman:2018dem,Chapman:2018lsv,
Moosa:2017yvt,Fan:2018xwf,Bernamonti:2019zyy,Swingle:2017zcd,Alishahiha:2018tep,An:2018xhv,Jiang:2018pfk,Kim:2017qrq,Yang:2019gce,Guo:2019vni,Cai:2016xho,
Lehner:2016vdi,Huang:2016fks,Cai:2017sjv,Cano:2018aqi,Jiang:2018sqj,Jiang:2019fpz,Feng:2018sqm,Alishahiha:2017hwg,
Carmi:2017jqz,Jiang:2019pgc,Liu:2019smx,Flory:2018akz,Flory:2019kah,Ghodrati:2018hss,Ghodrati:2017roz,Zhou:2019jlh}. However, there are some drawbacks for the two proposals as well. This motivates people to search new proposals for complexity \cite{Couch:2016exn,Fan:2018wnv,Fan:2019mbp}. In addition, the active research in holography promotes studies of complexity for quantum field theories \cite{JM1,HM1,RKK,CPMP,GHMR,CEHH,Bhattacharyya:2018bbv,JSY,Jiang:2018nzg,WFA,
Camargo:2018eof,Ali:2018fcz,CKMT,Bhattacharyya:2018wym,YANZ,Yang:2018tpo,Yang:2019udi} as well as in condensed matter physics \cite{Xiong:2019xoh}.

In this paper, we focus on the CA proposal. It states that the complexity of a holographic state is dual to the gravitational action evaluated on a certain region of spacetimes, called the Wheeler-DeWitt (WDW) patch, which is defined as the causal development of a bulk Cauchy surface that is anchored on the boundary state. One has
\begin{equation}
\mathcal{C}=\frac{I_{grav}}{\pi\hbar}\,,
\end{equation}
where $\hbar$ is the Planck constant which will be set to unity throughout this paper. By using this proposal, the time dependence of complexity for neutral and charged black holes in Einstein's gravity has been well studied in \cite{Carmi:2017jqz}. However, from the point of view of holography, it is of great interest to explore holographic applications of higher curvature gravitational theories, see for example \cite{Brigante:2007nu,Camanho:2009vw,Buchel:2009sk,deBoer:2009gx,Camanho:2009hu,Ge:2009ac,deBoer:2011wk,Chen:2013rcq,Dong:2013qoa}. The higher curvature corrections in the bulk, according to the spirit of AdS/CFT, are dual to finite N and finite coupling effects in the boundary CFTs and hence such bulk theories are dual to more general CFTs than those dual to the Einstein's gravity. In this paper, we would like to generalize the discussions of \cite{Carmi:2017jqz} and study the full time dependence of complexity for higher order gravities. In particular, to avoid ghost-like modes, we focus on Lovelock black holes in our numerical studies, despite that our discussions for action growth rate are valid to more general cases.

The remaining of this paper is organized as follows. In section 2, we briefly review the thermodynamics of Lovelock black holes. In section 3, under reasonable assumption, we deduce the action growth rate for neutral black holes in higher curvature theories. In section 4, we adopt the result in section 3 to numerically study the complexity growth rate for black holes in third order Lovelock gravities. In section 5, we further examine the complexity growth rate for charged Lovelock black holes. We conclude in section 6.

\section{Thermodynamics of Lovelock black holes: a brief review}
For later purpose, let's first review the thermodynamics of Lovelock black holes. The Lagrangian density for Lovelock gravities is given by
\be \mathcal{L}=R+(D-1)(D-2)\ell^{-2}+\sum_{n=2}^{[\fft{D-1}{2}]}\lambda_n\fft{(D-1-2n)!}{(D-3)!}(-1)^n \ell^{2n-2}\,\chi_{2n} \,,\ee
where $D$ is the spacetime dimensions, $\ell$ is the bare AdS radius, $\lambda_n$ are higher order gravitational coupling constants and $\chi_{2n}$ are Euler densities, give by
\be \chi_{2n}=\fft{(2n)!}{2^n}\,\delta^{[\mu_1\cdots \mu_n]}_{\nu_1\cdots\nu_n}R^{\nu_1\nu_2}_{\mu_1\mu_2}\cdots R^{\nu_{2n-1}\nu_{2n}}_{\mu_{2n-1}\mu_{2n}} \,.\ee
The static spherical/toric/hyperbolic solutions of the theories have been extensively studied in literature, for example
\cite{Cai:2001dz,Nojiri:2001aj,MyersSimon1988,Cai:2003kt,Dehghani:2009zzb}. One has
\be\label{ansatz} ds^2=-f(r)dt^2+\fft{dr^2}{f(r)}+r^2 d\Omega_{D-2\,,k}^2 \,,\ee
where $k=1\,,0\,,-1$ for which the $d\Omega_{D-2\,,k}^2$ is the metric for a $(D-2)$-dimensional sphere, torus and hyperboloid. The metric function $f(r)$ according to the field equations is determined by
\be\label{metricf} P\Big(\fft{\ell^2\big(f(r)-k \big)}{r^2} \Big)=\fft{16\pi G M\ell^2}{(D-2)\omega_{D-2}r^{D-1}} \,,\ee
where $M$ stands for the mass of the solutions and $\omega_{D-2}$ denotes the volume of the $(D-2)$-dimensional subspace. The function $P(x)$ is a polynomial, given by
\be\label{functionh} P(x)=1-x+ \sum_{n=2}^{[\fft{D-1}{2}]}\lambda_n x^n\,.\ee
Since we are particularly interested in planar black holes in this paper, we shall take $k=0$ at first in the following.

For a black  hole solution, the event horizon is defined by the largest root of the equation $f(r_h)=0$. Then using the relation (\ref{metricf}), one easily finds
\be\label{ntmass} M=\fft{(D-2)\omega_{D-2}r_h^{D-1}}{16\pi G\ell^2} \,.\ee
Moreover, evaluating the derivative with respect to $r$ for (\ref{metricf}) yields
\be f'(r_h)=\fft{(D-1)r_h}{\ell^2} \,.\ee
Thus, the temperature of the solutions is given by
\be T=\fft{f'(r_h)}{4\pi}=\fft{(D-1)r_h}{4\pi\ell^2} \,.\ee
On the other hand, the entropy of the solutions can be evaluated by using the Wald's entropy formula \cite{wald1,wald2} or the Jacobson-Myers' result \cite{Jacobson:1993xs}. Interestingly, it was found that the entropy for a planar Lovelock black hole is always equal to one quarter of the area of the event horizon \cite{Cai:2001dz,Nojiri:2001aj,MyersSimon1988,Cai:2003kt}
\be\label{entropy} S=\fft{\omega_{D-2}r_h^{D-2}}{4G} \,.\ee
Here it should be emphasized that the above results for planar black holes are valid to Lovelock theories with arbitrary densities since the thermodynamic quantities do not explicitly depend on any higher order coupling constant. On the contrary, for spherical/hyperbolic black holes, the thermodynamic quantities are given by
\bea
&&M=\fft{(D-2)\omega_{D-2}r_h^{D-1}}{16\pi G\ell^2}\,\Big[1+\fft{k\ell^2}{r^2_h}+\sum_{n=2}^{[\fft{D-1}{2}]}\lambda_n \Big(-\fft{k\ell^2}{r^2_h} \Big)^n \Big]\,,\nn\\
&&T=-\fft{k}{2\pi r_h}+\fft{(D-1)r_h}{4\pi\ell^2}\,\ft{1+\fft{k\ell^2}{r^2_h}+\sum_{n=2}^{[\fft{D-1}{2}]}\lambda_n \Big(-\fft{k\ell^2}{r^2_h} \Big)^n }{1-\sum_{n=2}^{[\fft{D-1}{2}]}n\lambda_n \Big(-\fft{k\ell^2}{r^2_h} \Big)^{n-1}}\,,\nn\\
&&S=\fft{\omega_{D-2}r_h^{D-2}}{4G}\,\Big[1-\sum_{n=2}^{[\fft{D-1}{2}]}\fft{n(D-2)}{D-2n}\lambda_n \Big(-\fft{k\ell^2}{r^2_h} \Big)^{n-1} \Big] \,,
\eea
which however heavily depend on the higher order coupling constants.

From the above results, it is straightforward to verify that the thermodynamical first law $dM=T dS$ holds. In addition, for planar black holes there exists a generalised Smarr formula $M=\fft{D-2}{D-1}T S$ which is associated to an extra scaling symmetry of the solutions\footnote{Under the scaling symmetry, the thermodynamic quantities behave as
$ M\rightarrow a^{D-1}M\,, T\rightarrow a T\,, S\rightarrow a^{D-2}S\,.$ The standard scaling dimensional arguments will lead to the generalised Smarr formula.}
\be\label{scalingsymmetry} r\rightarrow a r\,,\quad(t\,,x^i)\rightarrow a^{-1} (t\,,x^i)\,,\quad f\rightarrow a^2 f\,.\ee

It turns out that with a Maxwell field $\mathcal{L}_A=-\ft 14 F^2$, the metric for electrically charged solutions can still take the ansatz (\ref{ansatz}) whilst the gauge potential is solved as
\be A=\Phi\big(1-\ft{r_h^{D-3}}{r^{D-3}} \big)dt \,,\ee
where $\Phi$ is the chemical potential, given by ($Q$ is the electric charge)
\be \Phi=\fft{16\pi G Q}{(D-3)\omega_{D-2}r_h^{D-3}}\,,\quad Q=\fft{1}{16\pi G}\oint {}^*F \,.\ee
The metric function $f(r)$ now is determined by
\be\label{metricfc} P\Big(\fft{\ell^2\big(f(r)-k \big)}{r^2} \Big)=\fft{16\pi G M\ell^2}{(D-2)\omega_{D-2}r^{D-1}}-\fft{128\pi^2G^2Q^2\ell^2}{(D-2)(D-3)\omega^2_{D-2}r^{2(D-2)}} \,,\ee
where the function $P(x)$ is still given by (\ref{functionh}). Following the neutral case, it is straightforward to derive the mass and temperature for charged planar black holes. One has
\bea\label{MTcharge}
&&M=\fft{(D-2)\omega_{D-2}r_h^{D-1}}{16\pi G\ell^2}+\fft{8\pi G Q^2}{(D-3)\omega_{D-2}r_h^{D-3}}\,,\nn\\
&&T=\fft{(D-1)r_h}{4\pi\ell^2}-\fft{32\pi G^2Q^2}{(D-2)\omega_{D-2}^2r_h^{2D-5}}\,.
\eea
The entropy is still given by (\ref{entropy}). It follows that the first law $dM=T dS+\Phi dQ$ and the generalised Smarr formula $M=\ft{D-2}{D-1}\big(TS+\Phi Q \big)$ are satisfied.

In this paper, we are interested in the charged black holes with an inner horizon\footnote{In general, the existence of an inner horizon for charged black holes constrains the higher order coupling constants for Lovelock theories \cite{Cano:2018aqi}.}. We use the subscript ``+" (``-") to denote the thermodynamical quantities associated to the outer (inner) horizon. In this convention, one has
\be A=\Phi_+ \big(1-\ft{r_+^{D-3}}{r^{D-3}} \big)dt \,,\quad \Phi_\pm=\fft{16\pi G Q}{(D-3)\omega_{D-2}r_\pm^{D-3}} \,,\ee
where $\Phi_\pm$ are chemical potentials associated to the horizons, though $\Phi_-$ does not have a boundary dual. For later convenience, we would like to define the temperature on the inner horizon as $T_-=-T_+|_{r_+\longleftrightarrow r_-}$. Explicitly one has
\be T_\pm=\pm\fft{(D-1)r_\pm}{4\pi\ell^2}\mp\fft{32\pi G^2Q^2}{(D-2)\omega_{D-2}^2r_\pm^{2D-5}}\,. \ee
Moreover, a useful relation to connect the outer and inner horizon is
\bea\label{masshorizon}
M&=&\fft{(D-2)\omega_{D-2}r_+^{D-1}}{16\pi G\ell^2}+\fft{8\pi G Q^2}{(D-3)\omega_{D-2}r_+^{D-3}}\nn\\
&=&\fft{(D-2)\omega_{D-2}r_-^{D-1}}{16\pi G\ell^2}+\fft{8\pi G Q^2}{(D-3)\omega_{D-2}r_-^{D-3}}\,.
\eea
It is easy to check that the first law $dM=\pm T_\pm dS+\Phi_\pm dQ$ holds as well as the generalised Smarr formula $M=\ft{D-2}{D-1}\big(\pm T_\pm S_\pm+\Phi_\pm Q \big)$.

\section{Action growth rate for general higher order gravities}\label{sec2}
To study the full time dependence of complexity using the CA proposal, we shall first derive the action growth rate for general higher order gravities of the type
\be\label{theory} I_{bulk}=\fft{1}{16\pi G}\int_{\mathcal{M}} d^Dx \sqrt{-g}\,\mathcal{L}(g_{\mu\nu}; R_{\mu\nu\rho\sigma} )  \,.\ee
We consider generally static black hole solutions with spherical/hyperbolic/toric isometries
\be ds^2=-h(r)dt^2+\fft{dr^2}{f(r)}+r^2 d\Omega_{D-2\,,k}^2 \,.\ee
In general, the metric function $h(r)\neq f(r)$, except for certain theories (for example, the Lovelock gravities). Sometimes, it will be more convenient to perform the calculations using the ingoing and outgoing coordinates
\bea
u&=&t+r^*\,,\nn\\
v&=&t-r^*\,,
\eea
where the tortoise coordinate is defined as $r^*(r)=-\int^\infty_r \mathrm{d}r/w(r)f(r)$ with $w(r)=\sqrt{h(r)/f(r)}$. Notice that in our convention $r^*(\infty)=0$. The metric can be written as
\bea
ds^2=-h(r)du^2+2w(r) dudr+r^2d\Omega_{D-2\,,k}^2\,,
\eea
in the ingoing coordinate and
\bea
ds^2=-h(r)dv^2-2w(r) dvdr+r^2d\Omega_{D-2\,,k}^2\,,
\eea
in the outgoing coordinate, respectively.

It should be emphasized that for static black holes in higher curvature gravities, there may exist an alternative singularity located at a finite radii behind the event horizon. In this case, the WDW patch terminates at the new singularity and the action growth rate may become divergent\footnote{We find that at least, this is the case for various solutions in third order Lovelock theories, including Gauss-Bonnet black holes.}. Therefore, in the following, we will always assume the black hole solutions under consideration have only one singularity at the center of spacetimes. In general, this constrains the higher order gravitational coupling constants (for instance, for Gauss-Bonnet gravity (\ref{GBlagrangian}), the coupling constant $\lambda$ should be positive definite).
\begin{figure}[htbp]
  \centering
  \includegraphics[width=3.5in]{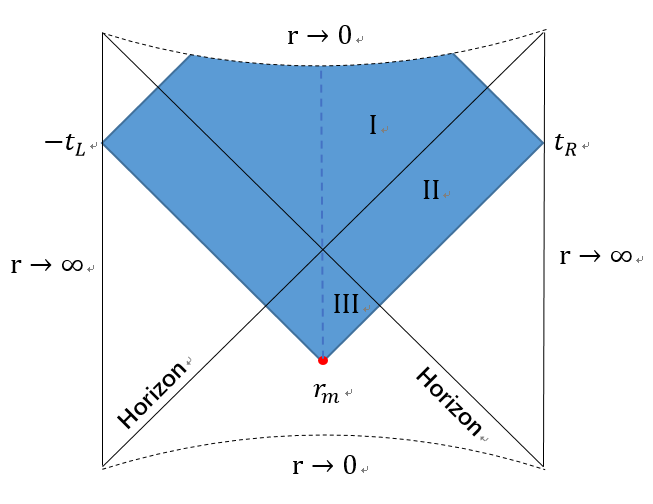}
  	\caption{The Wheeler-DeWitt (WDW) patch for a neutral two-sided AdS black hole. It moves forward in a symmetric way $t_L=t_R=t/2$. The dotted lines $r\rightarrow 0$ denote the locus of a spacelike singularity, where the WDW patch terminates. }
\label{tfd} \end{figure}

Under the above assumption, the Wheeler-DeWitt (WDW) patch for a neutral two-sided AdS black hole will be similar to that of a Schwarzschild black hole, as depicted in Fig.\ref{tfd}. In previous studies \cite{Mahapatra:2018gig,Carmi:2017jqz}, to calculate the action growth rate, one directly substitutes the solutions and the Lagrangian density into the various terms of the gravitational action. Despite that the results in \cite{Mahapatra:2018gig,Carmi:2017jqz} are valid, the discussions there are limited to certain cases. In this section, to keep our discussions as general as possible, we will extend the framework in \cite{Carmi:2017jqz} to general higher order gravities, without using the explicit form of the solutions as well as the Lagrangian density. Compared to \cite{Mahapatra:2018gig,Carmi:2017jqz}, it is more clear from our discussions how the various terms are cancelled in the gravitational action. This will help us to understand the final results better.

The total gravitational action evaluated on the WDW patch contains several different contributions, given by
\be\label{ntgravityaction} I_{grav}=I_{bulk}+I_{GHY}+I_{joint}+\cdots \,,\ee
where $I_{GHY}$ stands for the generalized Gibbons-Hawking-York boundary term evaluated at the future/past singularity and $I_{joint}$ stands for the action of the joints at which two null boundaries of the WDW patch intersects. The dots denotes the boundary terms that should be introduced on the future and past null boundaries \cite{Lehner:2016vdi}. Explicitly, one has \cite{Jiang:2018sqj}
\bea
&&I_{\kappa}=\fft{1}{2\pi }\int_{\mathcal{N}}d\lambda\mathrm{d}^{D-2}\theta\,\sqrt{|\gamma|}\,s\, \kappa\,,\nn\\
&&I_{ct}=\fft{1}{2\pi}\int_{\mathcal{N}} d\lambda d^{D-2}\theta \sqrt{|\gamma|}\,\hat{\Theta} \log{\big(\ell_{ct}\Theta \big)}\,,
\eea
where $I_\kappa$ is associated to parameterization of the outward-directed null norms $k^\mu$ of the null boundaries. One has
$k^\nu\nabla_\nu k^\mu=\kappa\,k^\mu$ and $s$ is the Wald entropy density (\ref{waldentropydensity}). Clearly the existence of this term breaks the reparameterization invariance of the gravity action. To eliminate this ambiguity, a counter term $I_{ct}$ is introduced in \cite{Lehner:2016vdi}. Here $\ell_{ct}$ is an arbitrary length scale, $\Theta $ is the scalar expansion defined by $\Theta=\nabla_\mu k^\mu$ and $\hat{\Theta}=\nabla_\mu\big(s k^\mu \big)$. Though the inclusion of this counter term is not necessary for the action principle, it plays an indispensable role in shockwave geometries \cite{Chapman:2018dem,Chapman:2018lsv,Moosa:2017yvt,Fan:2018xwf} as well as in other time dependent geometries \cite{Bernamonti:2019zyy}. However, for static black holes it does not have any time dependence and hence is irrelevant to current paper.

In the following, we will calculate the growth rate for the each term illustrated in (\ref{ntgravityaction}). The final result is presented in Eq.(\ref{totalaction}).

Before proceeding, we point out that the time evolution of complexity is essentially determined by the evolution of the joint $r_m$, which is characterized by
\be\label{positionrm} t\equiv t_R+t_L=-2r^*(r_m) \,.\ee
The result depends only on the combination of the two boundary times. This is a reminiscent of the symmetry of the problem. For later convenience, we always choose $t_R=t_L=t/2$. It turns out that there exists a critical time $t_c$, at which the joint $r_m$ lifts off of the past singularity, namely $t_c=-2r^*(0)$. When $t<t_c$, the action of the WDW patch remains a constant and hence $dI_{grav}/dt=0$. However, when $t>t_c$, the joint $r_m$ begins to grow with time. The growth rate is given by
\be\label{positionrm2} \fft{dr_m}{dt}=-\fft{w(r_m)f(r_m)}{2} \,.\ee
It follows that $dr_m/dt>0$ because of $f(r_m)<0$ since the position is in black holes interior. Thus, during the evolution the joint $r_m$ grows monotonically with the boundary time and approaches the event horizon from below at late times.

\subsection{Growth rate of bulk action}\label{bulkrate}
To calculate the bulk action and its growth rate, we split the WDW patch into three pieces (see Fig.\ref{tfd}) by following \cite{Carmi:2017jqz}. By simple calculations, one finds
\bea\label{fullbulkaction}
&&I_{\mathrm{I}}=\fft{\omega_{D-2}}{16\pi G}\int_{\epsilon}^{r_h}dr\,\sqrt{-\bar g}\,\mathcal{L}\,\big[t-2r^*(r) \big]\,,\nn\\
&&I_{\mathrm{II}}=-\fft{\omega_{D-2}}{16\pi G}\int_{r_h}^{\infty}dr\,\sqrt{-\bar g}\,\mathcal{L}\,4r^*(r)\,,\nn\\
&&I_{\mathrm{III}}=\fft{\omega_{D-2}}{16\pi G}\int_{r_m}^{r_h}dr\,\sqrt{-\bar g}\,\mathcal{L}\,\big[-t-2r^*(r) \big]\,,
\eea
where $\bar{g}=g/\gamma$, $\gamma$ is the metric determinant of $d\Omega_{D-2\,,k}^2$. Here and below, it was understood that $\epsilon\rightarrow 0$. Evaluating the derivative with respect to time, one finds
\bea \fft{dI_{bulk}}{dt}=\fft{\omega_{D-2}}{16\pi G}\int_{\epsilon}^{r_m}dr\sqrt{-\bar g}\,\mathcal{L}\,,
\eea
where we have adopted the relation (\ref{positionrm}).

 Without knowing the explicit form of the solutions as well as the Lagrangian density, it was established in \cite{Fan:2018wnv,Fan:2019mbp} that the bulk action can be expressed as several boundary terms via the Wald-Iyer Noether charge. One has
\be \nabla_\nu \mathcal{Q}^{\mu\nu}=j^\mu-\xi^\mu\mathcal{L} \,.\label{central2}\ee
where $\bm{\mathcal{Q}}$ is the 2-form Noether charge associated to the vector field $\xi$. For the general higher order gravities (\ref{theory}), it is given by \cite{Fan:2018qnt}
\be\label{generalNoether} \mathcal{Q}^{\mu\nu}=-2E^{\mu\nu\rho\sigma}\nabla_\rho \xi_\sigma-4\xi_\rho \nabla_\sigma E^{\mu\nu\rho\sigma}\,,\qquad E^{\mu\nu\rho\sigma}\equiv \fft{\partial \mathcal{L}}{\partial R_{\mu\nu\rho\sigma}}\,.\ee
$\bm j$ is the one-form presympletic current, which depends linearly on the variation of the dynamical fields (the metric and matter fields, collectively denoted by $\psi$). In particular, when $\xi$ is a Killing vector field, $j^\mu(\delta\psi)=0$ because of $\delta \psi=L_\xi \psi=0$. It follows that when $\xi=\partial/\partial t$, one has for static black holes
\be\label{bulk} \partial_r\Big(\sqrt{-g}\mathcal{Q}^{rt} \Big)=\sqrt{-g}\,\mathcal{L} \,.\ee
 Therefore, the bulk action defined on any space-time region can split into several boundary terms of the same region. Using the above identity, we obtain
 \bea\label{bulkaction}
 \fft{dI_{bulk}}{dt}=-\fft{\omega_{D-2}}{16\pi G}\,\sqrt{-\bar g}\mathcal{Q}^{tr}\,\Big|^{r_m}_{\epsilon}=-\hat{T}(r)\hat{S}(r)\Big|^{r_m}_{\epsilon}\,,
 \eea
where $\hat{T}(r)\,,\hat{S}(r)$ are temperature and Wald entropy functions defined at any $t=const\,,r=const$ hypersurface, given by
\be\label{TSfunction} \hat{T}(r)=\fft{h'(r)}{4\pi w(r)}\,,\quad \hat{S}(r)=-\fft{1}{8G}\int_{\Sigma_{D-2}}d^{D-2}y \sqrt{\gamma}\, \fft{\partial\mathcal{L}}{\partial R_{\mu\nu\rho\sigma}}\,\varepsilon_{\mu\nu}\varepsilon_{\rho\sigma}  \,,\ee
where $\varepsilon_{\mu\nu}$ is the binormal vector of the hypersurface. The two functions should not be confused with the temperature and entropy of black holes, which are defined at the event horizon. However, when the hypersurface approaches the event horizon at late times, they indeed reduce to the true temperature and entropy of black holes.

\subsection{GHY surface term }
For the bulk theories (\ref{theory}), the generalised Gibbons-Hawking-York (GHY) boundary term $I_{GHY}$ evaluated at a spacelike singularity is given by \cite{Deruelle:2009zk}
\be\label{GHY}  I_{GHY}=-\fft{1}{8\pi G}\int_{r=\epsilon}d^{D-1}x\sqrt{|h|}\,2\,\fft{\partial\mathcal{L}}{\partial R_{\mu\sigma\nu\rho}}\,n_\sigma n_\rho K_{\mu\nu} \,,\ee
where $n^\mu$ is the normal vector of a spacelike hypersurface $r=const$ and $K_{\mu\nu}$ is its second fundamental form. Notice that since this term is linearly proportional to the time lapse of the boundary, its growth rate is a constant. Without knowing more detail about the theories and the solutions, we cannot perform further calculations. However, we find that it is instructive to express the final result as follows
\be \fft{dI_{GHY}}{dt}=-\hat{T}(\epsilon)\hat{S}(\epsilon)+\Delta  \,,\ee
where the first term on the r.h.s will be cancelled by the bulk contributions whilst the second term $\Delta$, as will be shown later, is nothing else but the action growth rate at late times. Furthermore, from above relations, it is not hard to believe that the precise value of $\Delta$ (or its ratio to black hole mass) depends on the detail of the metric functions as well as the Lagrangian. In general, one does not expect it is universal to black holes in higher order gravities. We will show that this is indeed the case for Lovelock black holes. In particular, we find that its ratio to black hole mass is a characteristic constant, which is independent of higher order couplings. Hence, in the vanishing coupling limit, the result in general does not reduce to that of Schwarzschild black holes. While this may go against one's intuition, we find that the reason is natural from mathematical point of view. We will explain it in the next section.

\subsection{Growth rate of joint action}
The joint action was derived from action principle for gravitational theories defined on spacetimes with nonsmooth boundaries \cite{Lehner:2016vdi}. Recently, the result was extended to higher order gravities in \cite{Jiang:2018sqj,Cano:2018ckq}. One has
\be I_{joint}\equiv \fft{1}{2\pi}\int_{C_{D-2}}d\Omega_{D-2} \,s \,a \,,\ee
where $s$ is the Wald entropy density function
\be\label{waldentropydensity} s=-\fft{1}{8G}\, \fft{\partial\mathcal{L}}{\partial R_{\mu\nu\rho\sigma}}\,\varepsilon_{\mu\nu}\varepsilon_{\rho\sigma} \,,\ee
and $a$ is the standard corner term for Einstein's gravity \cite{Lehner:2016vdi}. For null-null joint, one has
\be a\equiv \eta\log{|\ft 12 k_1\cdot k_2|}\,,\qquad \eta=-\mathrm{sign}(k_1\cdot k_2)\,\mathrm{sign}(\hat{k}_1\cdot k_2) \,,\ee
where $k_i$ are outward directed normal vectors of the dual null boundaries and $\hat{k}_i$ are auxiliary null vectors defined in the tangent space of the boundaries, orthogonal to the joint and pointing outward from the boundary regions. For the codimension-2 hypersurfaces defined by $t=\mathrm{const}\,,r=\mathrm{const}$, one has
\be I_{joint}=\fft{1}{2\pi}\,\hat{S}(r)a(r) \,.\ee
The null normals can be written as
\bea
k_\mu&=&-\alpha\partial_\mu v=-\alpha\partial_\mu(t-r^{\ast})\nn\\
\bar{k}_\mu&=&\bar{\alpha}\partial_\mu u=\bar \alpha\partial_\mu (t+r^{\ast})
\eea
where $\alpha$ and $\bar{\alpha}$ are arbitrary positive constants which can be fixed by implementing the asymptotic normalizations $k\cdot \hat{t}_L=-\alpha$ and $\bar{k}\cdot\hat{t}_R=-\bar{\alpha}$, where $\hat{t}_{L,R}$ are the asymptotic Killing vectors on the left and right boundaries, respectively. Without loss of generality, we choose $\alpha=\bar\alpha$ so that
\bea
a=-\log\Big(\frac{|f(r_m)|}{\alpha^2}\Big)\,.
\eea
Evaluating the derivative of the joint action with respect to the boundary time yields
\bea \fft{dI_{joint}}{dt}&=&\hat{T}(r_m)\hat{S}(r_m)-\fft{\hat{S}'(r_m)}{4\pi} |f(r_m)|\log{\Big(\fft{|f(r_m)|}{\alpha^2} \Big)}\,.\eea
Note that the first term on the r.h.s will cancel a piece of bulk contributions (\ref{bulkaction}).
\subsection{Complexity growth rate at late times}\label{ntlatetime}
Combining all the results above together, we deduce
\be\label{totalaction} \fft{dI_{grav}}{dt}=\Delta-\fft{\hat{S}'(r_m)}{4\pi} |f(r_m)|\log{\Big(\fft{|f(r_m)|}{\alpha^2} \Big)} \,.\ee
It is immediately seen that the time evolution of the action (and complexity) is essentially determined by the evolution of the past joint $r_m$. At late times, $r_m$ approaches the event horizon $r_h$ and hence the second term on the r.h.s vanishes. Thus,
\be \fft{dI_{grav}}{dt}\Big|_{t\rightarrow \infty}=\Delta \,,\ee
implying that the rate of change of complexity at late times is given by
\be  \fft{d\mathcal{C}}{dt}\Big|_{t\rightarrow \infty}=\fft{\Delta}{\pi} \,.\ee
To study the behavior of complexity around late times more carefully, we shall include its next-to-leading order term. In this case, the position $r_m$ behaves as
\be\label{rmlate} r_m=r_h\big(1-c_m\, e^{-2\pi T t}+\cdots\big) \,,\ee
where $c_m$ is a positive constant, specified by
\be c_m=2\,\mathrm{exp}\Big[ F(r_h)\int_{r_h}^\infty H(r)dr \Big]\,.\ee
Here the functions $F(r)$ and $H(r)$ are introduced in (\ref{ntF}) and (\ref{ntH}).

By plugging the result (\ref{rmlate}) into (\ref{totalaction}), one finds
\be\label{cplate} \fft{d\mathcal{C}}{dt}=\fft{\Delta}{\pi}+2c_m\,T r_h\hat{S}'(r_h)\, Tt\, e^{-2\pi T t}+\cdots \,.\ee
 Interestingly, the next-to-leading order term is always positive definite. Therefore, for the CA proposal, the complexity growth rate generally approaches the late time limit from above. Consequently, it will violate any conjectured upper bound given by the growth rate at late times \cite{Brown:2015bva,Brown:2015lvg,Cai:2016xho}.

\subsection{Numerical approach}
 Now we are ready to study the full time dependence of complexity for neutral Lovelock black holes. In general, we shall adopt  a numerical approach to solve the joint $r_m$ as a function of $t$ via (\ref{positionrm}) at first and then substitute the result into
 (\ref{totalaction}) to obtain the rate of change of complexity. In general, this is not a hard problem but one should notice that the tortoise coordinate behaves singular at the event horizon. To cure this issue, we introduce a new function $F(r)$ as\footnote{Notice that for Lovelock black holes $w(r)=1$. However, our discussions are valid to black holes with $w(r)\neq 1$ as well, with the function $F(r)$ redefined as $w(r)f(r)\equiv F(r)(r^2-r_h^2)$. }
\be\label{ntF} f(r)\equiv F(r)(r^2-r_h^2) \,.\ee
The inverse of the metric function $f(r)$ can be written as
\be \fft{1}{f(r)}=\fft{1}{4\pi T}\Big(\fft{1}{r-r_h}-\fft{1}{r+r_h}+F(r_h)H(r) \Big)\,,\ee
where we have adopted a useful relation $F(r_h)=2\pi T/r_h$ and the function $H(r)$ is defined by
\be\label{ntH} H(r)\equiv \fft{2r_h \big( F(r_h)-F(r)\big)}{F(r)F(r_h)(r^2-r_h^2)}\,.\ee
It is easy to see that $H(r)$ is a regular function at the event horizon. Then the tortoise coordinate can be integrated to
\be r^*(r)=\fft{1}{4\pi T}\log{\Big|\fft{r-r_h}{r+r_h}\Big|}-\fft{1}{2r_h}\int_r^\infty H(\tilde r)d\tilde r \,.\ee
Since the singular part has been isolated, it is straightforward to solve the tortoise coordinate numerically.

In the subsequent sections, we will calculate the rate of change of complexity for various black hole solutions in third order Lovelock theories (including Gauss-Bonnet black holes) by using the framework illustrated above. Without confusion, we focus on planar black holes in the remaining of this paper.

\section{The time dependence of complexity for Gauss-Bonnet black holes}\label{EGBcomp}

Let us first study a simpler case: the time dependence of complexity for Gauss-Bonnet black holes. The Lagrangian density is given by
\be\label{GBlagrangian} \mathcal{L}=R-2\Lambda+\ft{\lambda\,\ell^{2}}{(D-3)(D-4)}\big(R^2-4R^2_{\mu\nu}+R^2_{\mu\nu\lambda\rho} \big) \,,\ee
where $\Lambda=-\ft 12(D-1)(D-2)\ell^{-2}$ is the bare cosmological constant. The black holes exist in $D\geq 5$ dimensions, given by
\bea\label{GBBH}
&&ds^2=-f(r)dt^2+\fft{dr^2}{f(r)}+r^2 dx^i dx^i\,, \nn\\
&&f(r)=\fft{r^2}{2\lambda\ell^2}\Big[1-\sqrt{1-4\lambda+\ft{64\pi\lambda\ell^2 G M}{(D-2)\omega_{D-2}r^{D-1}}}\, \Big]\,.
\eea
 From the metric function $f(r)$, it is immediately seen that for a negative $\lambda$, the solution will have an alternative singularity at a finite radii behind the event horizon. To avoid this, we demand  $\lambda>0$, which is also a physically interesting case from string theory perspective. It was established in \cite{Brigante:2007nu,Camanho:2009vw} that the Gauss-Bonnet coupling was strongly constrained by microcausality (or positive energy fluxes) of the boundary theories. The allowed region for the coupling constant is \cite{Brigante:2007nu,Camanho:2009vw}
\be\label{lambda} 0< \lambda\leq \ft{(D-3)(D-4)(D^2-3D+8)}{4(D^2-5D+10)^2} \,.\ee

For Gauss-Bonnet gravity, the Noether charge is given by \cite{wald2,Fan:2014ala,Chen:2016qks}
 \bea\label{EGBNoether}
\mathcal{Q}^{\mu\nu} = -2\Big(\nabla^{[\mu}\xi^{\nu]}
+\ft{2\lambda\,\ell^{2}}{(D-3)(D-4)}\,\big(R\,\nabla^{[\mu}\xi^{\nu]}-4 R^{\sigma[\mu}\nabla_\sigma \xi^{\nu]}+ R^{\mu\nu\sigma\rho}\nabla_\sigma \xi_\rho \big)\Big)\,.
\eea
By definition, the Wald entropy function $\hat{S}(r_m)$ can be evaluated as
\be\label{entropyfunction} \hat{S}(r_m)=\ft{\omega_{D-2} r_m^{D-2}}{4G}\Big(1-\ft{2(D-2)\lambda\ell^2f(r_m)}{(D-4)r_m^2} \Big) \,.\ee
%It is straightforward to verify that the relation (\ref{bulkaction}) is indeed satisfied.
Evaluating the generalised GHY boundary term (\ref{GHY}) yields\footnote{A different GHY boundary term for GB gravity is given in \cite{Davis:2002gn}, where
\bea
I_{GHY}&=&-\fft{1}{8\pi G}\int_{\partial \mathcal{M}}\mathrm{d}^{D-1}x\sqrt{-h}\,\big(K+\ft{2\lambda\,\ell^{2}}{(D-3)(D-4)}\,J \big)\,,\nn\\
J&=&\fft 23 K^3-2K K_{\mu\nu}K^{\mu\nu}+\fft 43 K_{\mu\lambda}K^{\lambda\rho}K_\rho^{\,\,\,\mu}    \nn\\
&&+R_{\mu\lambda\nu\rho}h^{\mu\nu}h^{\lambda\rho} K -2R_{\mu\lambda\nu\rho}h^{\mu\nu}K^{\lambda\rho}    \nn  \,.
\eea
It leads to a different growth rate of complexity. In particular, at late times, it gives $d\mathcal{C}/dt=4M/3\pi$, which is valid to $D\geq 5$ dimensional solutions.}
\be
\Delta=\lim_{\epsilon\rightarrow 0}\ft{(D-2)\omega_{D-2}}{8(D-4)\pi G}\epsilon^{D-5}f(\epsilon)\Big[-(D-4)\epsilon^2+2\lambda\ell^2\Big((D-4)f(\epsilon)+\epsilon f'(\epsilon)\Big)\Big]\,.
\ee
Substituting the metric function $f(r)$ into the equation, one finds
\be\label{GBlate} \Delta=\fft{2(D-3)M}{D-4} \,.\ee
It is interesting to note that the ratio $\Delta/M$ is a pure number, independent of the higher order coupling constant $\lambda$. In fact, this is a generic feature for Lovelock black holes.

A puzzle immediately appears. While the action and the solution smoothly reduce to those of Einstein's gravity in the $\lambda\rightarrow 0$ limit, the action growth rate does not reduce to the known value $\Delta=2M$ for Schwarzschild black holes. One may argue that this is unreasonable \cite{Cai:2016xho}. However, we are aware of that the result can be attributed to a simple fact: different orders of the limits $\epsilon\rightarrow 0$ and $\lambda\rightarrow 0$ lead to different results, namely
\be\label{lateinequality} \lim_{\lambda\rightarrow 0}\lim_{\epsilon\rightarrow 0} \Delta(\epsilon\,,\lambda)\neq \lim_{\epsilon\rightarrow 0}\lim_{\lambda\rightarrow 0} \Delta(\epsilon\,,\lambda) \,.\ee
From this perspective, the result is mathematically sound. Furthermore, according to (\ref{totalaction}), for GB black holes the ratio of the action growth rate to the mass in the $\lambda\rightarrow 0$ limit should differ from that of Schwarzschild black holes by a fixed constant at any time $t>t_c$ in the evolution. We will show this explicitly in our numerical results.
\begin{figure}[htbp]
  \centering
  \includegraphics[width=210pt]{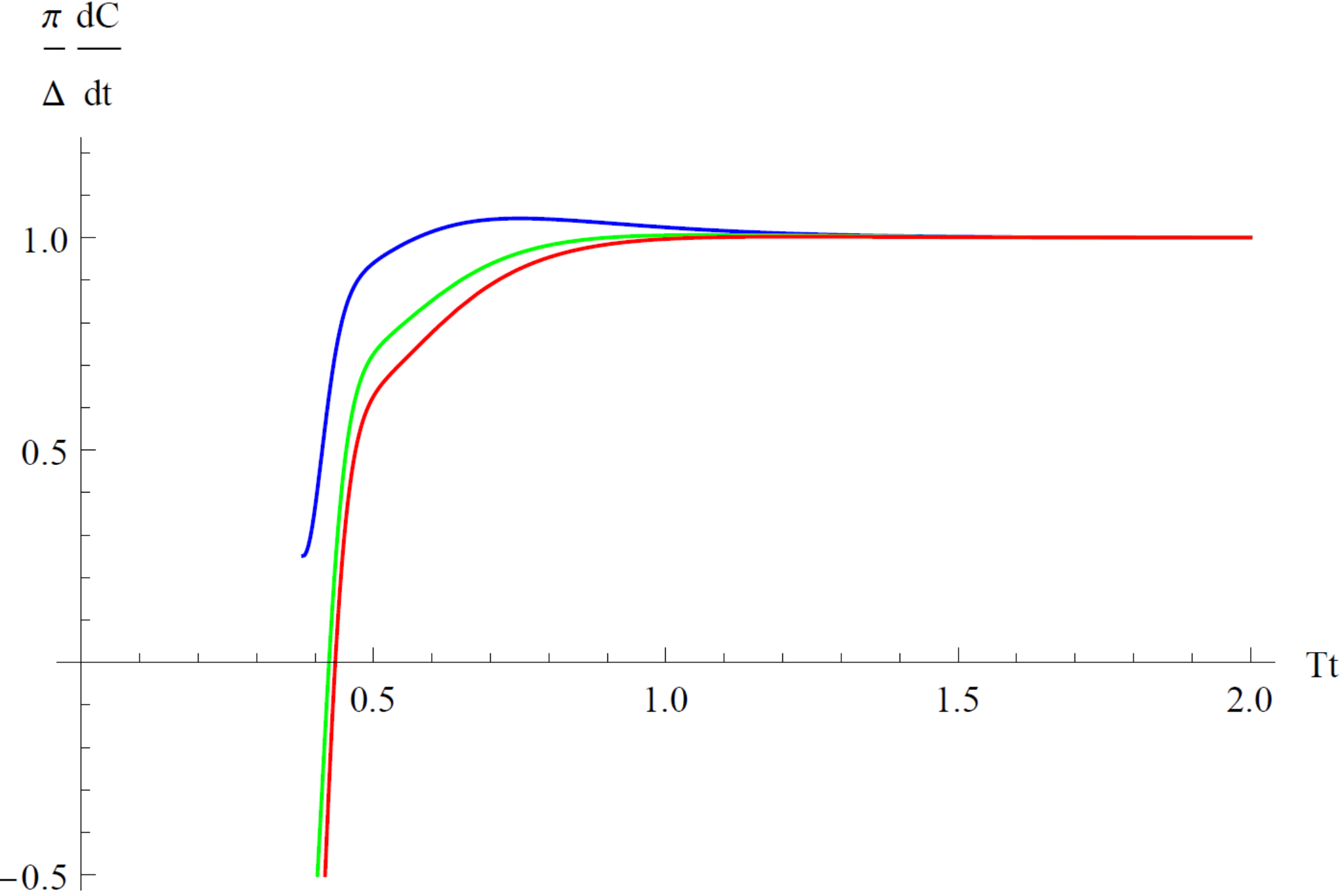}
  \includegraphics[width=210pt]{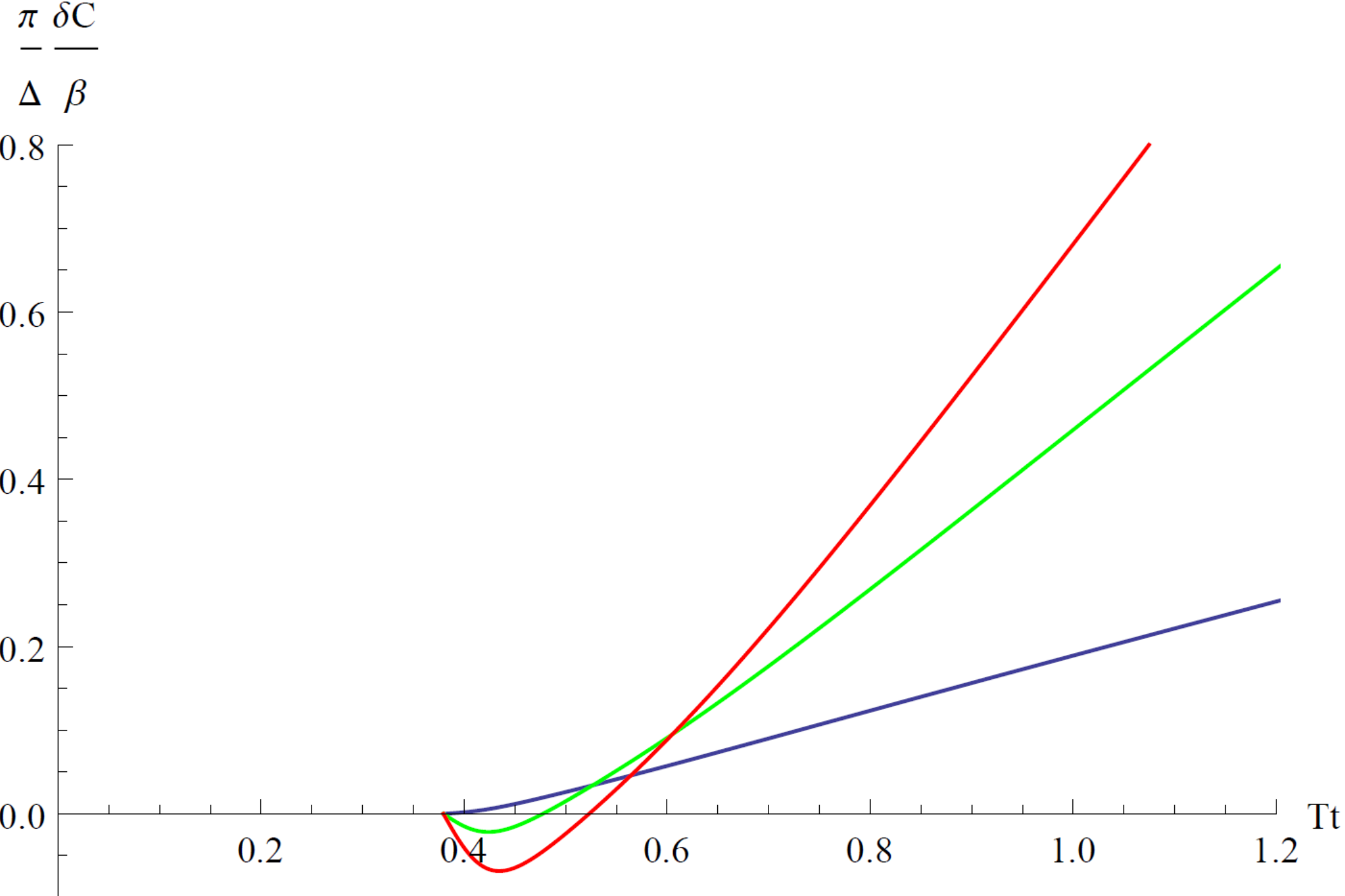}
  	\caption{Left panel: complexity growth rate for different sizes of $D=5$ dimensional Gauss-Bonnet black holes with $\lambda=0.05$. Right panel: the complexity difference $\delta \mathcal{C}=\mathcal{C}(t)-\mathcal{C}(t_c)$ found by integrating $d\mathcal{C}/dt$. The horizon radius are chosen as: $r_h/\ell=1$ (Blue), $r_h/\ell=3$ (Green) and $r_h/\ell=5$ (Red). We have set $G=\alpha=1\,,\omega_3=16\pi$.}
\label{GB5} \end{figure}
%For a black hole with a given $\lambda$, the evolution of complexity (normalized by its late time rate) as a function of $T t$ is identical for the planar solution at different temperatures. This is %closely related to the fact that for AdS planar black holes, there exists an extra scaling symmetry $r\rightarrow a r\,,(t\,,x^i)\rightarrow a^{-1}(t\,,x^i)\,,f(r)\rightarrow a^2 f(r)$, which implies %$T\rightarrow a T\,, S\rightarrow a^{D-2} S$. It is clear that the dimensionless time $T t$ is invariant under the global scaling. Furthermore, in spite of that the complexity growth rate scales as %$\ft{d\mathcal{C}}{dt}\rightarrow a \ft{d\mathcal{C}}{dt}$, the normalized rate $\fft{\pi \hbar}{\Delta}\ft{d\mathcal{C}}{dt}$ is also invariant under the scaling.

With the above results in hand, we are ready to numerically compute the rate of change of complexity for GB black holes. In Fig.\ref{GB5}, we show the complexity growth rate as a function of the dimensionless time $t/\beta=Tt$ for different sizes of $D=5$ dimensional GB black holes with a same coupling constant $\lambda=0.05$ as well as the complexity difference $\delta \mathcal{C}=\mathcal{C}(t)-\mathcal{C}(t_c)$ found by integrating $d\mathcal{C}/dt$. It is easy to see that the critical time measured by the thermal time $\beta=1/T$ is identical for different sizes of black holes while the complexity and its growth rate behave significantly different. To explain this, we work in the dimensionless coordinate $z=r/r_h$ and introduce two functions $\tilde{f}(z)$ and $\tilde{S}(z)$ as follows
\be f(r)\equiv \fft{r_h^2}{\ell^2}\,\tilde{f}(z)\,,\qquad \hat{S}(r)\equiv r_h^{D-2}\tilde{S}(z) \,.\ee
Notice that these new functions are scaling invariant under the global scaling symmetry (\ref{scalingsymmetry}). Explicitly, for GB black holes one has
\bea
&&\tilde{f}(z)=\fft{z^2}{2\lambda}\Big(1-\sqrt{1-4\lambda+4\lambda \,z^{1-D}} \,\Big)\,,\nn\\
&&\tilde{S}(z)=\ft{\omega_{D-2} z^{D-2}}{4G}\Big(1-\ft{2(D-2)\lambda\tilde{f}(z)}{(D-4)z^2} \Big)\,.
\eea
By plugging the above results into (\ref{totalaction}), we deduce
\bea
\fft{\pi}{\Delta}\fft{d\mathcal{C}}{dt}&=&1+\ft{r_h^{D-1}}{4\pi\Delta\ell^2}\,\tilde{S}'(z)\tilde{f}(z)\log{|\ft{r_h^2}{\ell^2}\tilde{f}(z)|}\nn\\
&=&1+\ft{4G M}{(D-2)\omega_{D-2}\Delta }\,\tilde{S}'(z)\tilde{f}(z)\log{|\ft{r_h^2}{\ell^2}\tilde{f}(z)|}\,,\eea
where in the second equality we have adopted the thermodynamical relation (\ref{ntmass}) for Lovelock black holes. This implies that though the normalized growth rate of complexity is dimensionless, it scales nontrivially under the global scaling symmetry (\ref{scalingsymmetry}). We find
\be \fft{\pi}{\Delta}\fft{d\mathcal{C}}{dt}\rightarrow \fft{\pi}{\Delta}\fft{d\mathcal{C}}{dt}+\ft{8G M}{(D-2)\omega_{D-2}\Delta }\,\tilde{S}'(z)\tilde{f}(z)\log{\big(\fft{a}{\ell}\big)} \,.\ee
This explains why and how the complexity and its growth rate depend on black hole sizes.

In addition, using the dimensionless coordinate, the critical time can be expressed as
\be t_c=-2r^*(0)=\fft{D-1}{2\pi T}\int_0^{\infty}\fft{dz}{\tilde{f}(z)} \,.\ee
Since $Tt_c$ scales invariant under the global scaling symmetry (\ref{scalingsymmetry}), it should be identical for different sizes of black holes with same higher order couplings.

It is worth emphasizing that the above discussions do not depend on the explicit expression of $\tilde{f}(z)$ and hence are valid to Lovelock black holes in higher order theories.

\begin{figure}[htbp]
  \centering
  \includegraphics[width=230pt]{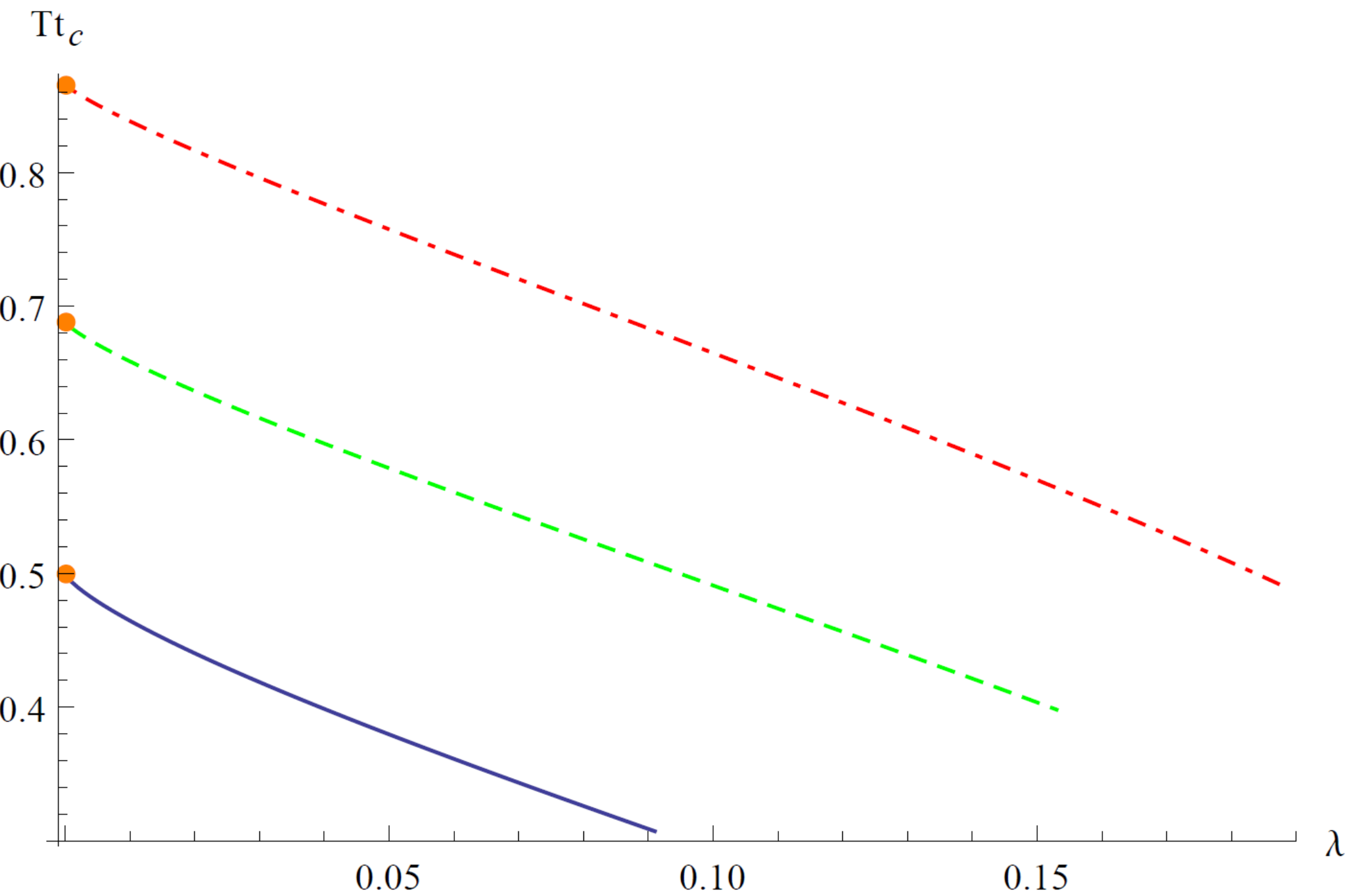}
  \caption{The critical time measured by $1/T$ as a function of the Gauss-Bonnet coupling for $D=5$ (blue solid), $D=6$ (green dashed) and $D=7$ (red dotdashed) dimensional black holes, respectively. The orange points denote the value of Schwarzschild black holes in the same dimensions. }\label{tcGB}
\end{figure}

In Fig.\ref{tcGB}, we show $Tt_c$ as a function of $\lambda$ for $D=5\,,6\,,7$ dimensional solutions. We find that it is always a decreasing function of the GB coupling and approaches the value of Schwarzschild black holes (in the same dimensions) in the vanishing coupling limit.

\begin{figure}[htbp]
  \centering
  \includegraphics[width=210pt]{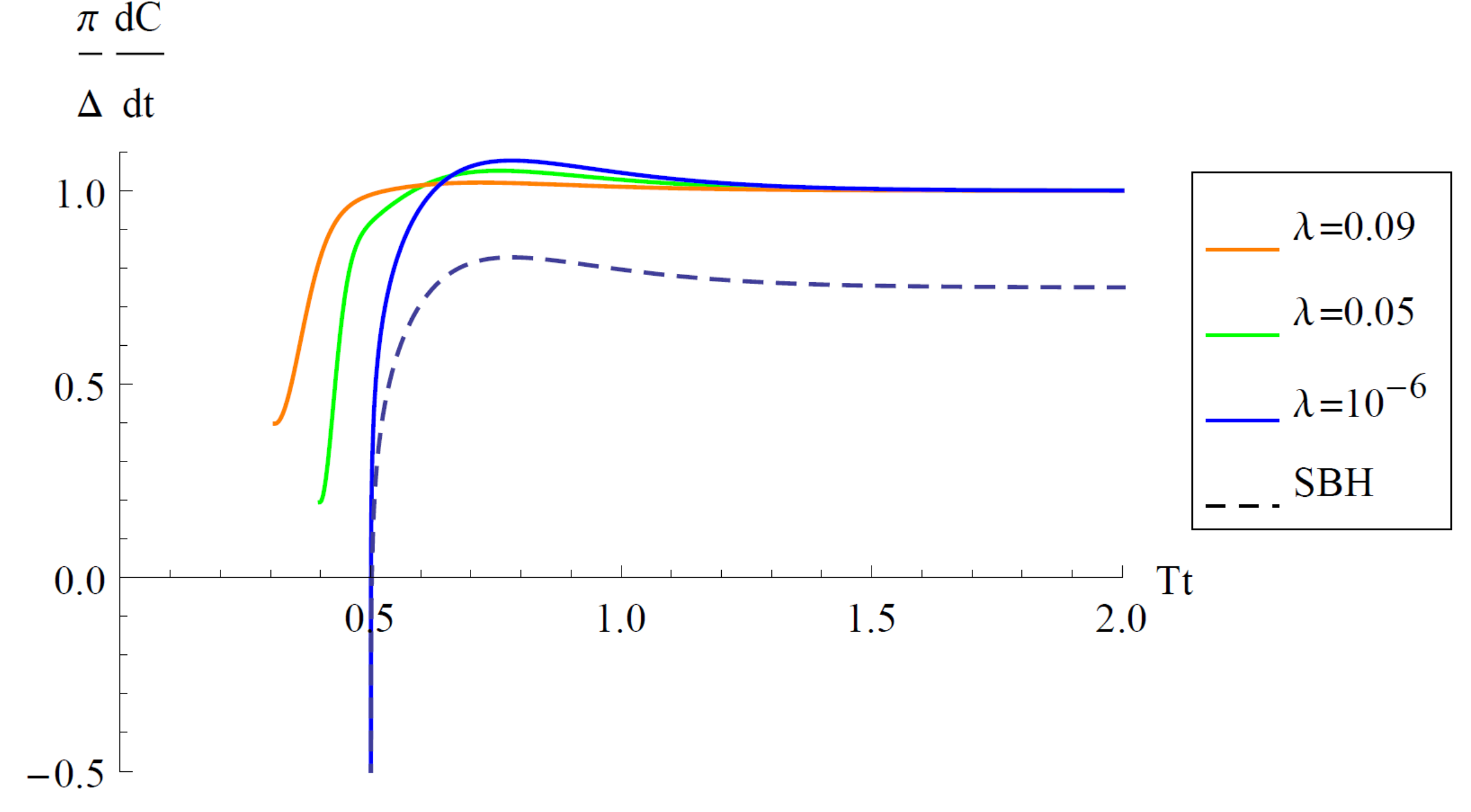}
  \includegraphics[width=210pt]{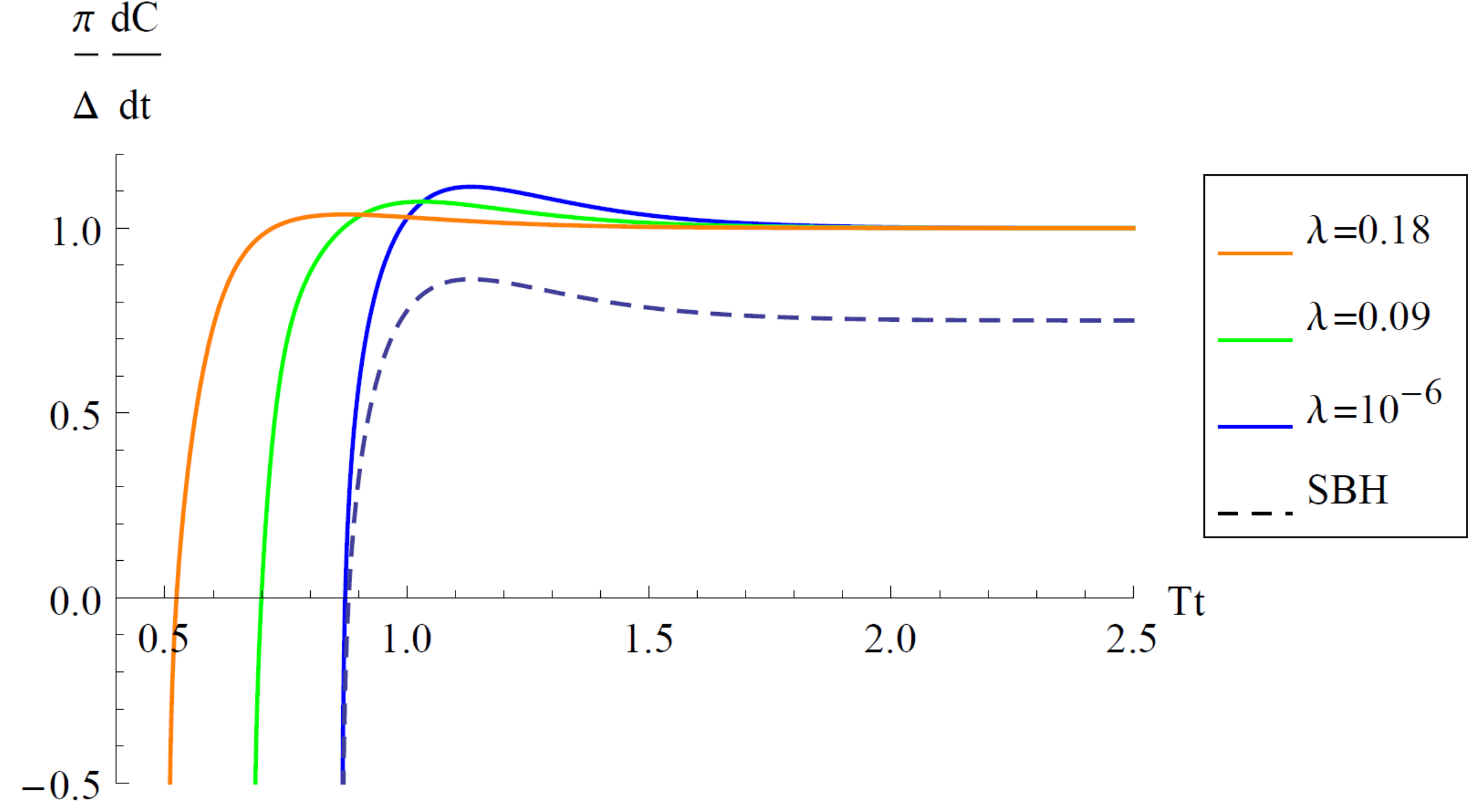}
  	\caption{The time derivative of complexity for $D=5$ (left) and $D=7$ (right) dimensional Gauss-Bonnet black holes with various couplings in the causal region. The dashed lines stand for the Schwarzschild results which have been moved along the vertical axis properly. We have set $r_h/\ell=1\,,G=\alpha=1\,, \omega_{D-2}=16\pi$. }
\label{GB} \end{figure}

In Fig.\ref{GB}, we show the numerical results for the time derivative of complexity for $D=5$ and $D=7$ dimensional Gauss-Bonnet black holes with various couplings $\lambda$ in the causal region as a function of the dimensionless time $Tt$. We have chosen $r_h/\ell=1$ for convenience. Clearly, the rate of change of complexity always approaches a local maximum before arriving at the late time limit. This is consistent with our half-analytical result (\ref{cplate}). Moreover, the maximum value becomes bigger for a smaller $\lambda$. For a sufficiently small $\lambda$, the result is roughly same as that of Schwarzschild black holes, up to a fixed constant. For example, for $\lambda=10^{-6}$, our numerical result nearly coincides with the Schwarzschild case by moving the latter along the vertical axis properly, as expected.

\textbf{Early times:}
In Fig.\ref{GB}, we also observe that the time derivative of complexity behaves very different for $D=5$ and $D=7$ dimensional black holes at very early times after $t_c$. To clarify this, we notice that the joint $r_m$ at early times is very close to the past singularity, namely $r_m\rightarrow 0$ as $t\rightarrow t_c$. In this case, the metric function $f(r)$ behaves as
\be f(r_m)=-c_1\Big(\fft{r_h}{r_m}\Big)^{\fft{D-5}{2}}+\fft{r_m^2}{2\lambda \ell^2}+O\big(r_m^{\fft{D+3}{2}}\big)\,,\quad c_1=\fft{r_h^2}{\sqrt{\lambda}\ell^2} \,.\ee
Substituting this into (\ref{positionrm2}), one finds at leading order
\be\label{rmearly} r_m=c_2\,r_h\Big(T(t-t_c) \Big)^{\fft{2}{D-3}}+\cdots \,,\quad c_2=\Big(\ft{(D-3)\pi}{(D-1)\sqrt{\lambda}}\Big)^{\fft{2}{D-3}} \,.\ee
It is straightforward to derive the complexity growth rate at leading order. It follows that for the $D=5$ dimensional solution, one has
\bea\label{cpearly1} \fft{\pi}{\Delta}\fft{d\mathcal{C}}{dt}
&=&1+\ft 14\log{\lambda}-\ft 14\log{\big( \ft{16\pi G M}{3\omega_3\ell^2} \big)}+O\Big(T^2(t-t_c)^2 \Big) \nn\\
&=&1+\ft 14\log{\lambda}-\log{\big(\fft{r_h}{\ell} \big)}+O\Big(T^2(t-t_c)^2 \Big)
\,.\eea
At leading order, the growth rate is a constant, which is an increasing function of $\lambda$. This matches with the left panel of Fig.\ref{GB}. However, for higher dimensional solutions $D\geq 6$, the complexity growth rate behaves logarithmical to leading order
\be\label{cpearly2} \fft{\pi}{\Delta}\fft{d\mathcal{C}}{dt}=\ft{D-5}{2(D-3)}\log{\Big(T(t-t_c) \Big)}+O\Big((t-t_c)^{0} \Big) \,.\ee
We find that it is perfectly matched with our numerical results in a variety of dimensions at very early times.

\section{The time dependence of complexity for third order Lovelock black holes}\label{LL3comp}
We continue studying the time evolution of complexity for black holes in third order Lovelock gravities. The Lagrangian density is given by
\be\label{LL3lagrangiantot}\mathcal{L}=R-2\Lambda+\ft{\lambda\,\ell^{2}}{(D-3)(D-4)}\big(R^2-4R^2_{\mu\nu}+R^2_{\mu\nu\lambda\rho} \big)+\ft{\mu\,\ell^4}{3(D-3)(D-4)(D-5)(D-6)}\,\mathcal{L}_{3} \,,\ee
%\be\label{LL3lagrangiantot}\mathcal{L}_{tot}=R+30\ell^{-2}+\ft{\lambda\,\ell^{2}}{12}\big(R^2-4R^2_{\mu\nu}+R^2_{\mu\nu\lambda\rho} \big)+\ft{\mu\,\ell^4}{72}\mathcal{L}_{3} \,,\ee
where $\mu$ is the third order coupling constant and
\bea\label{LL3lagrangian}
\mathcal{L}_3&=&R^3+3R R^{\mu\nu\alpha\beta}R_{\alpha\beta\mu\nu}-12R R^{\mu\nu}R_{\mu\nu}+24R^{\mu\nu\alpha\beta}R_{\mu\alpha}R_{\nu\beta}+16R^{\mu\nu}R_{\nu}^{\,\,\,\alpha}R_{\mu\alpha}\nn\\
&&+24R^{\mu\nu\alpha\beta}R_{\alpha\beta\nu\rho}R_{\mu}^{\,\,\,\rho}
+8R^{\mu\nu}_{\,\,\,\,\,\,\alpha\rho}R^{\alpha\beta}_{\,\,\,\,\,\,\nu\sigma} R^{\rho\sigma}_{\,\,\,\,\,\,\mu\beta}+2R_{\alpha\beta\rho\sigma}R^{\mu\nu\alpha\beta}R^{\rho\sigma}_{\,\,\,\,\,\, \mu\nu}\,.
\eea
%The black hole solution can be formally written as
%\be\label{LL3BH} ds^2=-f(r)dt^2+\fft{dr^2}{f(r)}+r^2 \sum_{i=1}^{5}dx^i dx^i\,,\quad f(r)=\ft{\lambda r^2}{\mu \ell^2}\Big(1-\varphi(r) \Big)  \,,\ee
%where the form of the function $\varphi(r)$ is complicated for a generic $\mu$ (see for example \cite{deBoer:2009gx,Camanho:2009hu}).
With a nonvanishing $\mu$, nontrivial black hole solutions exist in $D\geq 7$ dimensions. For simplicity, we focus on the $D=7$ dimensional solution in this section.

\begin{figure}[ht]
\centering
\includegraphics[width=300pt]{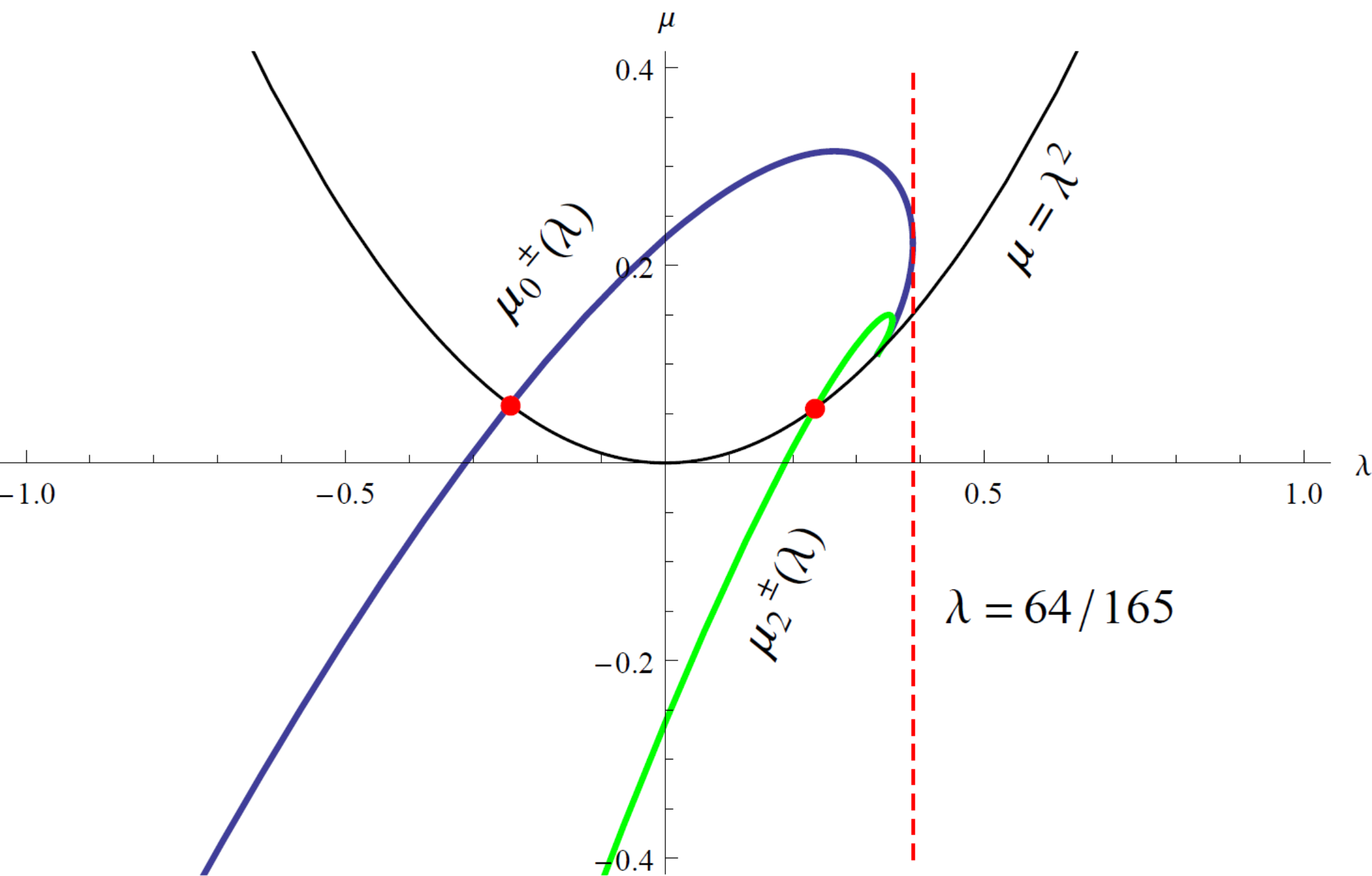}
\caption{{\it The higher order coupling constants $(\lambda\,,\mu)$ are strongly constrained by microcausality of the theories. The allowed parameters are described by the striped region between the blue curve and the green curve. The black curve is a special case $\mu=\lambda^2$ and the interval between the red points corresponds to $-91/375\leq\lambda\leq 19/81$.  } }
\label{causality}\end{figure}
It turns out that the new coupling constant $\mu$ strongly affects the causal structure of the theories. This is well studied in \cite{deBoer:2009gx,Camanho:2009hu}. The allowed region for the higher order coupling constants $(\lambda\,,\mu)$ is specified as
\be \mu_2^\pm(\lambda)\leq \mu\leq \mu_0^\pm (\lambda)\,,\ee
where
\bea
&&\mu_2^\pm(\lambda)=\fft{1}{243}\Big(189\lambda-32\pm 4(9\lambda-2)\sqrt{16-45\lambda} \Big)\,,\nn\\
&&\mu_0^\pm(\lambda)=\fft{1}{1125}\Big(315\lambda+128\pm 4(15\lambda+4)\sqrt{64-165\lambda} \Big)
\,,\eea
follow causality constraints for helicity-two and helicity-zero excitations, respectively. In Fig.\ref{causality}, the causal region  corresponds to the strip between the blue curve and the green curve . However, as the GB case, to avoid an alternative singularity behind the event horizon, we need further constrain the higher order coupling constants. This will be more clear when the metric function is given for certain cases. For example, for $\lambda=0$ the coupling constant $\mu$ should be positive definite, as can be seen from the metric function (\ref{LL3BH1}). This gives rise to $0<\mu\leq \ft{256}{1125}$. Another interesting case is $\mu=\lambda^2$. From the metric function (\ref{LL3BH2}), a negative $\lambda$ should be excluded and the allowed region becomes $0<\lambda\leq 19/81$, which is changed, compared to that of the $D=7$ dimensional GB gravity.

To proceed, we need evaluate the Noether charge, the Wald entropy function and the generalised GHY surface term associated to the third order Lovelock theories. To achieve this goal, we derive the tensor $E^{\mu\nu\rho\sigma}=\partial\mathcal{L}/\partial R^{\mu\nu\rho\sigma}$ in the Appendix A. The result is lengthy. We suggest the readers moving there for detail. It follows that for the $D=7$ dimensional black holes, the Wald function entropy is given by
\be \hat{S}(r_m)=\fft{\omega_5\, r_m^5}{4G}\Big(1-\ft{10\lambda \ell^2 f(r_m)}{3r_m^2}+\ft{5\mu \ell^4 f(r_m)^2}{r_m^4} \Big) \,. \ee
On the other hand, evaluating GHY surface term yields
\be\label{Lovelocklate} \Delta=\lim_{\epsilon\rightarrow 0}-\fft{5\omega_5}{24\pi G}\,f(\epsilon)
\Big[3\epsilon^4-2\lambda\ell^2\epsilon^2\Big(3f(\epsilon)+\epsilon\, f'(\epsilon) \Big)+3\mu\ell^4f(\epsilon)\Big( f(\epsilon)+2\epsilon\, f'(\epsilon)\Big)  \Big]\,.\ee
In the following, we will adopt these formulas to study the complexity growth rate for several solutions in third order Lovelock gravities.

\subsection{Einstein's gravity extended with a single third order Lovelock density}
To examine the effect of the new coupling constant $\mu$ on the time evolution of complexity, we turn off the Gauss-Bonnet term at first and consider the remaining theory: Einstein's gravity extended with a single third order density. For later convenience, we set $\mu=3/\big(2\tilde{\mu}^3\big)$, where $\tilde\mu \geq 15/8$. The black hole solution is given by
\bea\label{LL3BH1} &&ds^2=-f(r)dt^2+\fft{dr^2}{f(r)}+r^2 \sum_{i=1}^{5}dx^i dx^i\,,\nn\\
&& f(r)=\fft{\tilde\mu r^2}{3\ell^2}\Big(2\tilde\mu\, \phi(r)^{-1}-3\phi(r) \Big)  \,,\eea
where
\be \phi(r)=\Big(\sqrt{\ft{8\tilde{\mu}^3}{27}+\big(1-\ft{r_h^6}{r^6} \big)^2}-\big(1-\ft{r_h^6}{r^6} \big) \Big)^{1/3} \,.\ee
Substituting the metric function into (\ref{Lovelocklate}), we find that the action growth rate at late times is $\Delta=6M$. Again, the ratio $\Delta/M$ is a constant, independent of the higher order coupling $\mu$. Since $\Delta\neq 2M$, we expect that during the evolution, the normalized growth rate of complexity will differ from that of Schwarzschild black holes by a fixed constant in the limit $\mu\rightarrow 0$, as will be shown later.
\begin{figure}[htbp]
  \centering
  \includegraphics[width=230pt]{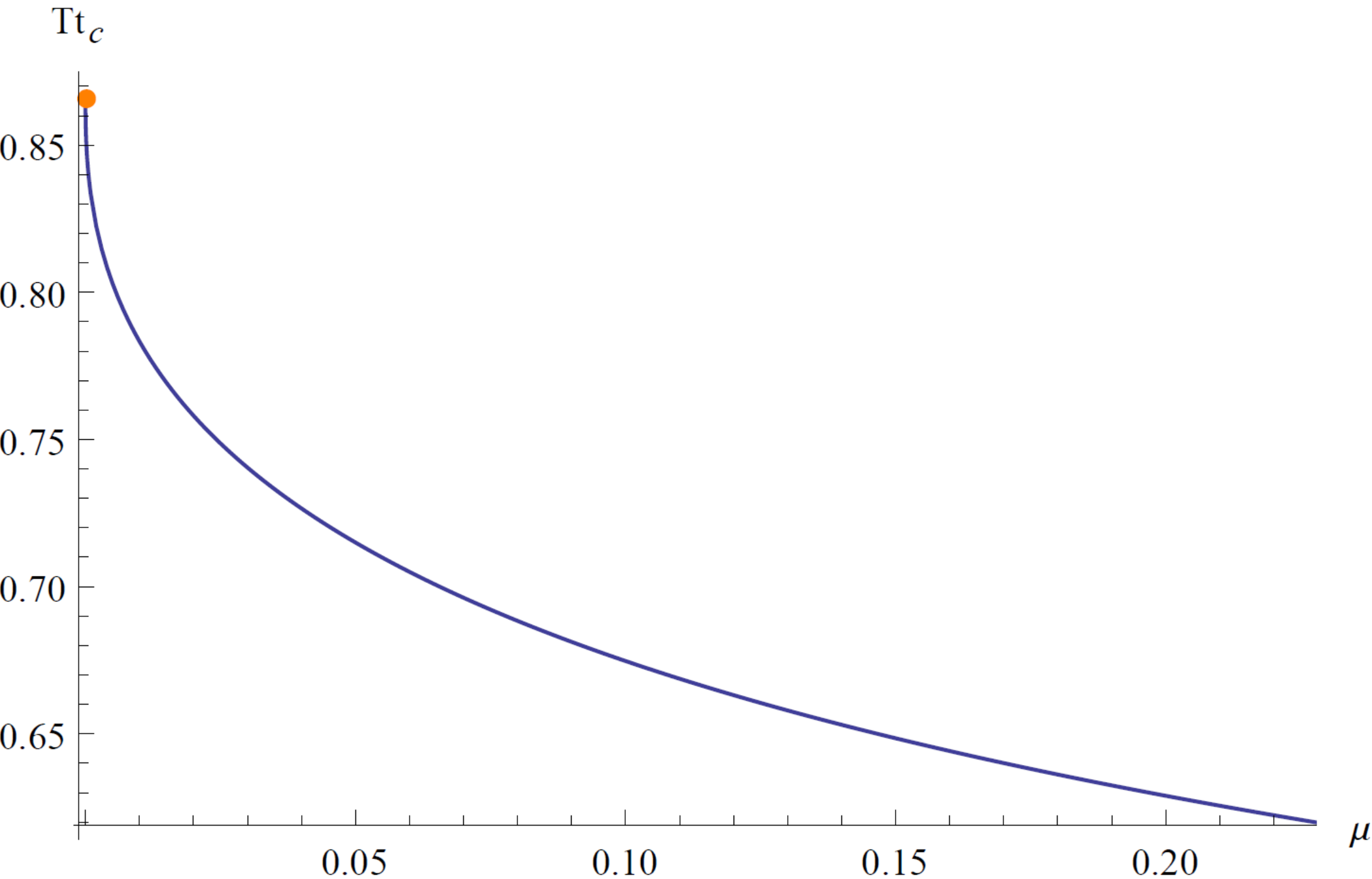}
  \caption{The critical time measured by the thermal time $1/T$ is plotted as a function of $\mu$. The orange point denotes the value of $D=7$ dimensional Schwarzschild black holes.}\label{tcLL3}
\end{figure}

By numerically solving the critical time $t_c$, we find that $t_c/\beta$ is a decreasing function of the higher order coupling constant $\mu$, similar to the Gauss-Bonnet case, as depicted in Fig.\ref{tcLL3}. For vanishing $\mu$, it approaches the value of $D=7$ dimensional Schwarzschild black holes.

It turns out that at very early times after $t_c$, the position $r_m$ grows linearly with time at leading order
\be\label{rmLL3} r_m=\big(9\mu \big)^{-1/3}\,\pi r_h\, T(t-t_c)+\cdots \,,\ee
so that the complexity growth rate behaves as
\bea\label{cpLL3s1} \fft{\pi}{\Delta}\fft{d\mathcal{C}}{dt}&=&1+\fft 16\log{\big(\fft{\mu}{3}\big)}-\fft 16\log{\big( \ft{16\pi G M}{5\omega_5\ell^4} \big)}+O\Big(T^4(t-t_c)^4 \Big)\nn\\
&=&1+\fft 16 \log{\big(\fft{\mu}{3}\big)}-\log{\big(\fft{r_h}{\ell} \big)}+O\Big(T^4(t-t_c)^4 \Big)\,.\eea
Interestingly, the result at leading order is a constant, depending on the logarithm of the higher order coupling $\mu$ and the black hole mass. This is very similar to the result of $D=5$ dimensional GB black hole (\ref{cpearly1}).
\begin{figure}[htbp]
  \centering
  \includegraphics[width=250pt]{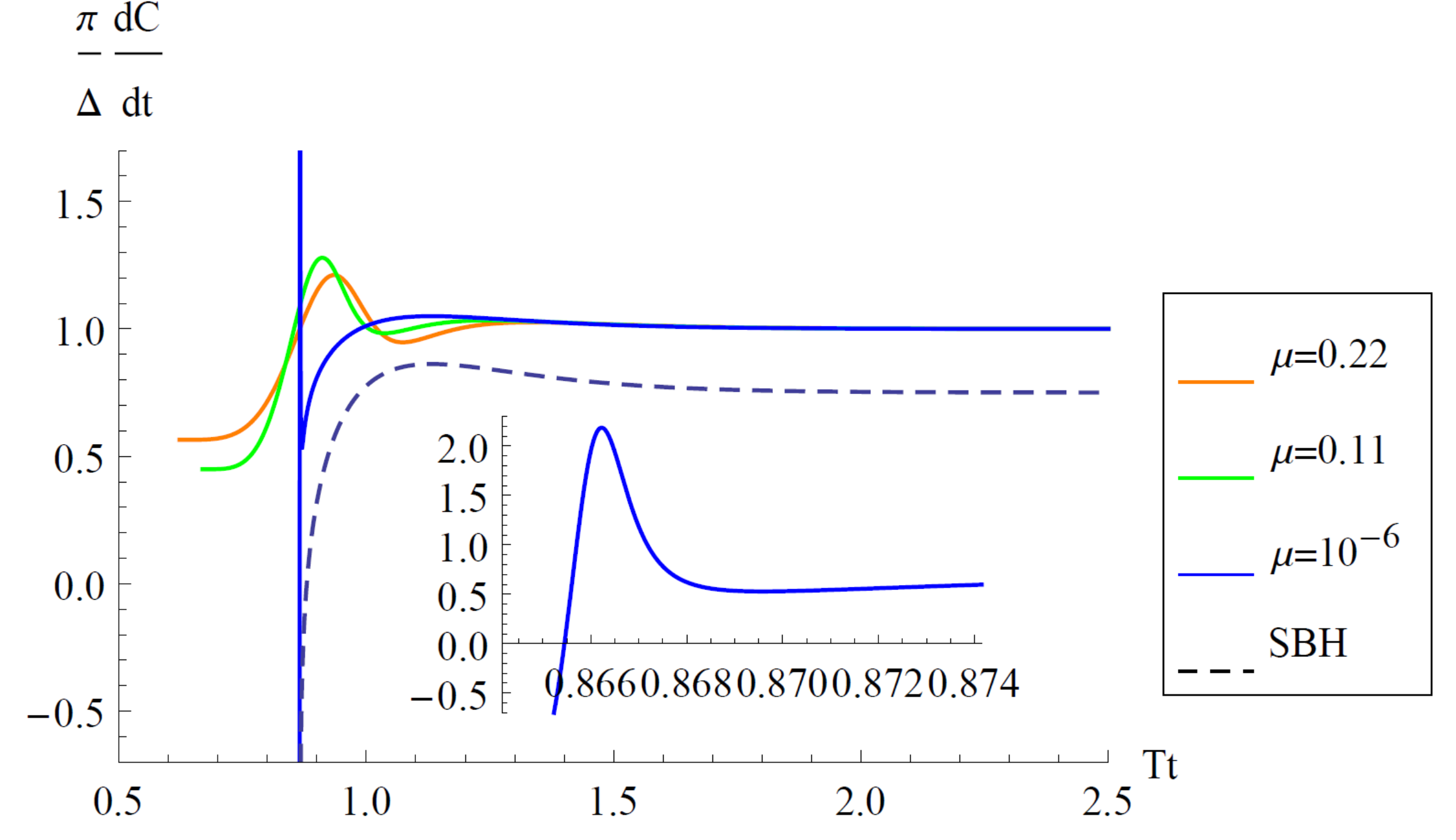}
  	\caption{The time derivative of complexity for the Lovelock black hole (\ref{LL3BH1}) with various couplings $\mu$ in the causal region as a function of the dimensionless time $Tt$. The dashed line denotes the result of a $D=7$ dimensional Schwarzschild black hole. We have set $G=\alpha=r_h=\ell=1\,, \omega_5=16\pi$. }
\label{LL31} \end{figure}

In Fig.\ref{LL31}, we show the full numerical result for the time derivative of complexity for various couplings $\mu$ in the allowed region as a function of the dimensionless time $Tt$. Indeed, at very early times the complexity growth rate is a constant, which increases as $\mu$ increases. After the early times, the growth rate increases with time until arriving at a maximum, which should be distinguished from the local maximum close to the late times. The latter is universal to all neutral black holes, as analyzed in sec.\ref{ntlatetime}. In fact, we find that the existence of this new maximum is peculiar to all the black holes in third order Lovelock gravities, except the reduced case $\mu=0$. Furthermore, it is interesting to note that while the height of the maximum increases as $\mu$ decreases, its width is gradually suppressed. For sufficiently small $\mu$, the peak will be smoothed and the time derivative of complexity will approach that of a Schwarzschild black hole, up to a fixed constant.

\subsection{A special case: $\mu=\lambda^2$}
With inclusion of the Gauss-Bonnet term, the general black hole solution becomes much more involved. Here we would like to consider a special case $\mu=\lambda^2$ at first. The solution greatly simplifies to
\bea\label{LL3BH2} &&ds^2=-f(r)dt^2+\fft{dr^2}{f(r)}+r^2 \sum_{i=1}^{5}dx^i dx^i\,,\nn\\
&& f(r)=\fft{ r^2}{\lambda \ell^2}\Big[1-\Big(1-3\lambda\big(1-\ft{r_h^6}{r^6} \big) \Big)^{1/3} \Big]  \,.\eea
Notice that the new coupling constant $\mu$ strongly effects the causal structure of the theory so that the allowed regime for $\lambda$ is changed to $ 0< \lambda\leq \fft{19}{81}$ \cite{deBoer:2009gx,Camanho:2009hu}. Here a negative $\lambda$ is excluded since it introduces an alternative singularity behind the event horizon.
\begin{figure}[htbp]
  \centering
    \includegraphics[width=210pt]{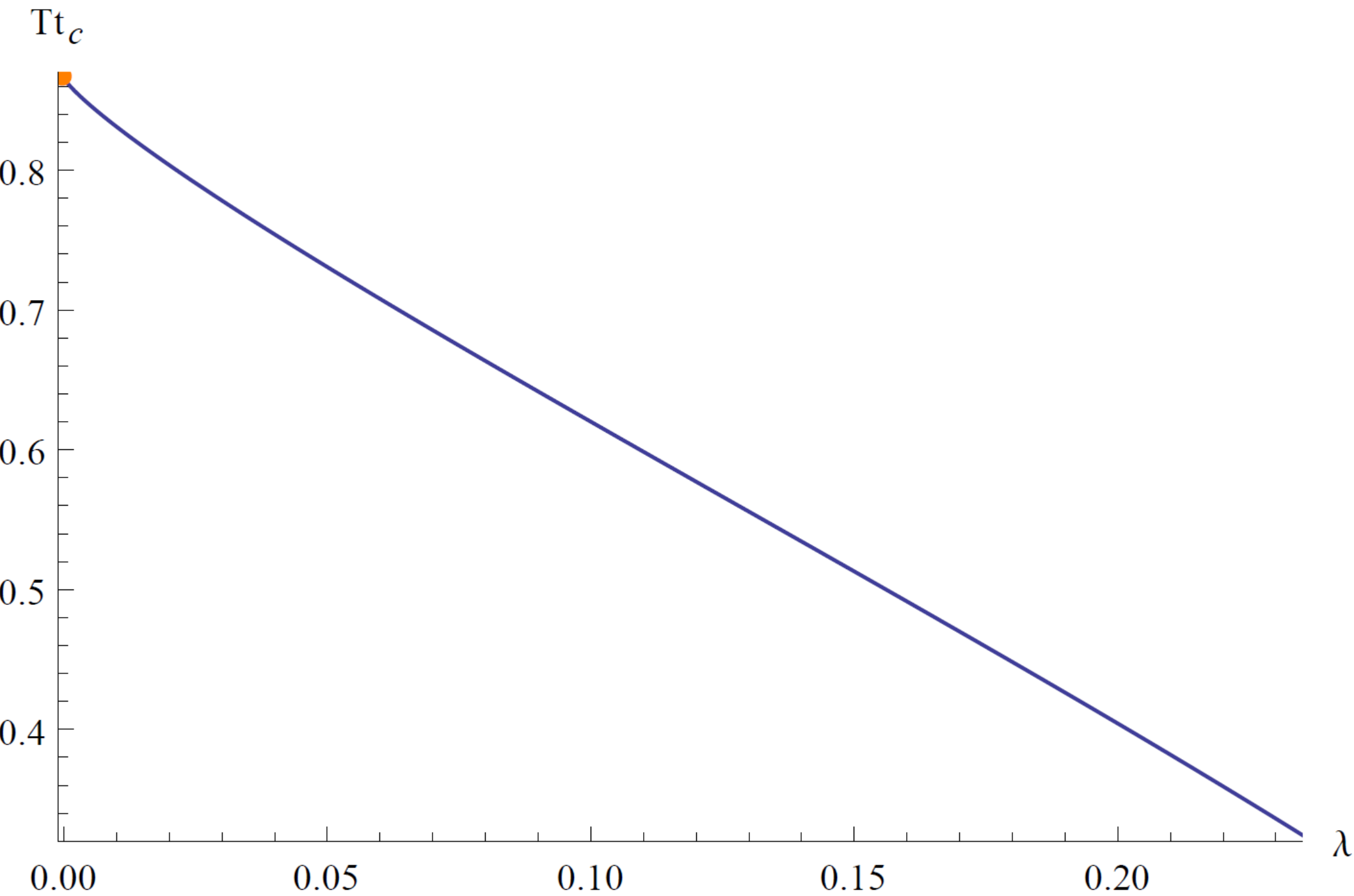}
    \includegraphics[width=220pt]{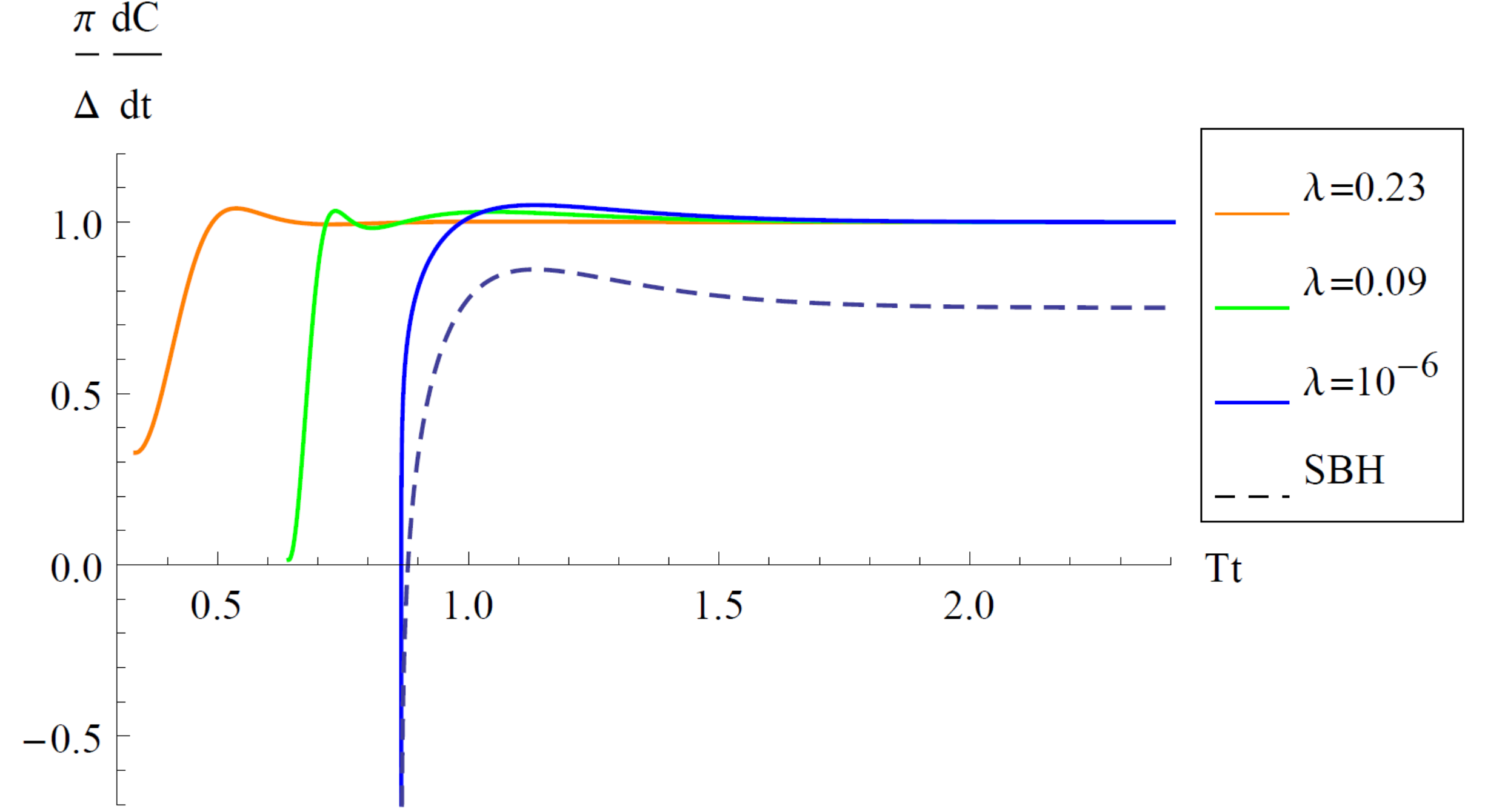}
  	\caption{ Left panel: the critical time $Tt_c$ is plotted as a function of $\lambda$. The orange point denotes the value of $D=7$ dimensional Schwarzschild black holes. Right panel: the evolution of complexity for the Lovelock black hole (\ref{LL3BH2}) with various couplings $\lambda$ in the causal region as a function of the dimensionless time $Tt$. The dashed line denotes the result of a $D=7$ dimensional Schwarzschild black hole. We have set $G=\alpha=r_h=\ell=1\,, \omega_5=16\pi$. }
\label{LL32} \end{figure}

Substituting the metric function into (\ref{Lovelocklate}), we obtain $\Delta=6M$, which is the same as the previous solution (\ref{LL3BH1}). In fact, they have more similar features:\\
$\bullet$ The critical time $Tt_c$ is a decreasing function of the higher order coupling $\lambda$, as shown in the left panel of Fig.\ref{LL32};\\
$\bullet$ At very early times, the joint $r_m$ grows linearly with time at leading order
\be r_m=\big(3\lambda \big)^{-2/3}\,\pi r_h\, T(t-t_c)+\cdots \,.\ee
As a consequence, the complexity growth rate behaves as
\be \fft{\pi}{\Delta}\fft{d\mathcal{C}}{dt}=1+\fft 16 \log{\big(\fft{\lambda^2}{3}\big)}-\log{\big(\fft{r_h}{\ell} \big)}+O\Big(T^2(t-t_c)^2 \Big)\,,\ee
where again the leading order is a constant, depending on the logarithm of the higher order coupling and the black hole mass.\\
$\bullet$ During the evolution, the complexity growth rate approaches a new maximum after the early times, as depicted in the right panel of Fig.\ref{LL32}. Its feature is similar to that of the solution (\ref{LL3BH1}). However, an important difference is it is more strongly suppressed for smaller couplings so that the complexity growth rate approaches the Schwarzschild result faster.

\subsection{Generic case}

For generic coupling constants $(\lambda\,,\mu)$, the black hole solution is given by \cite{Dehghani:2009zzb}
\bea\label{genericLLBH}
&&ds^2=-f(r)dt^2+\fft{dr^2}{f(r)}+r^2 \sum_{i=1}^{5}dx^i dx^i\,,\nn\\
&&f(r)=\fft{\lambda\, r^2}{\mu\ell^2}\Big[1+\Big(J(r)+\sqrt{\Omega(r)} \Big)^{1/3}-\Gamma \Big(J(r)+\sqrt{\Omega(r)} \Big)^{-1/3} \,\Big] \,,
\eea
where
\bea &&\Omega(r)=J(r)^2+\Gamma^3\,,\quad \Gamma=\fft{\mu}{\lambda^2}-1\,,\nn\\
&&J(r)=1-\fft{3\mu}{2\lambda^2}+\fft{3\mu^2}{2\lambda^3}\Big(1-\fft{r_h^6}{r^6} \Big) \,.\eea
Notice that for $\mu\neq \lambda^2$, there exists only one singularity at the center of spacetimes. Evaluating (\ref{Lovelocklate}), we find that the action growth rate at late times is
\be
\Delta=\left\{\begin{array}{ll}
0\,,\qquad\,\,\,\,\,\mathrm{for}\quad \lambda>0\,,\nn\\
6M\,,\qquad \mathrm{for}\quad \lambda<0\,.
\end{array}
\right.
\ee
However, the result for $\lambda>0$ is not physically sound. In fact, in this case the metric function behaves singular close to the singularity so that the complexity growth rate is not well behaved at early times.

In the following, we shall focus on the $\lambda<0$ case. The parameters relevant to an AdS black hole is $\mu>\lambda^2$, corresponding to the region enclosed by the blue curve and black curve in the second quadrant in Fig.\ref{causality}.

\begin{figure}[htbp]
  \centering
  \includegraphics[width=210pt]{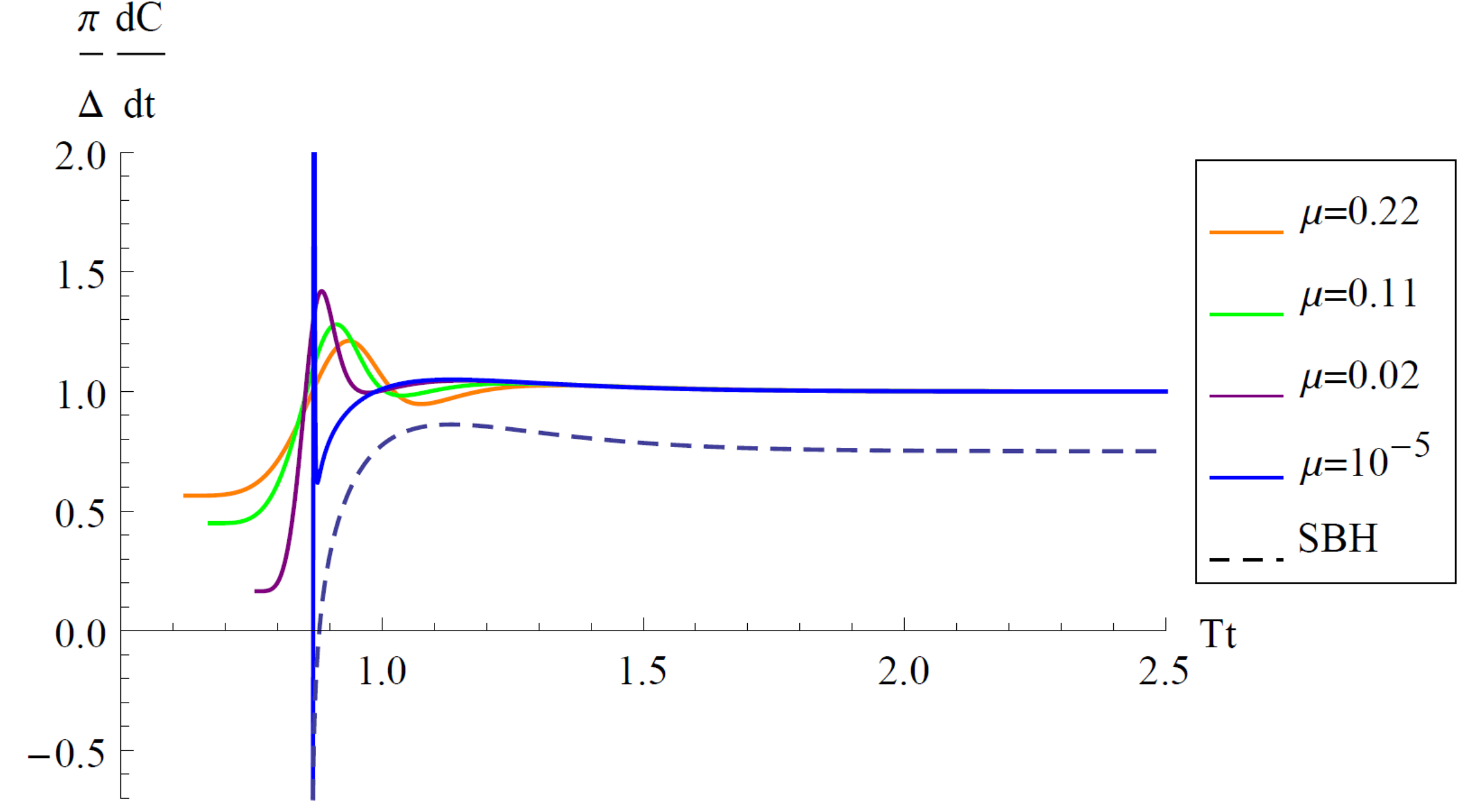}
  \includegraphics[width=210pt]{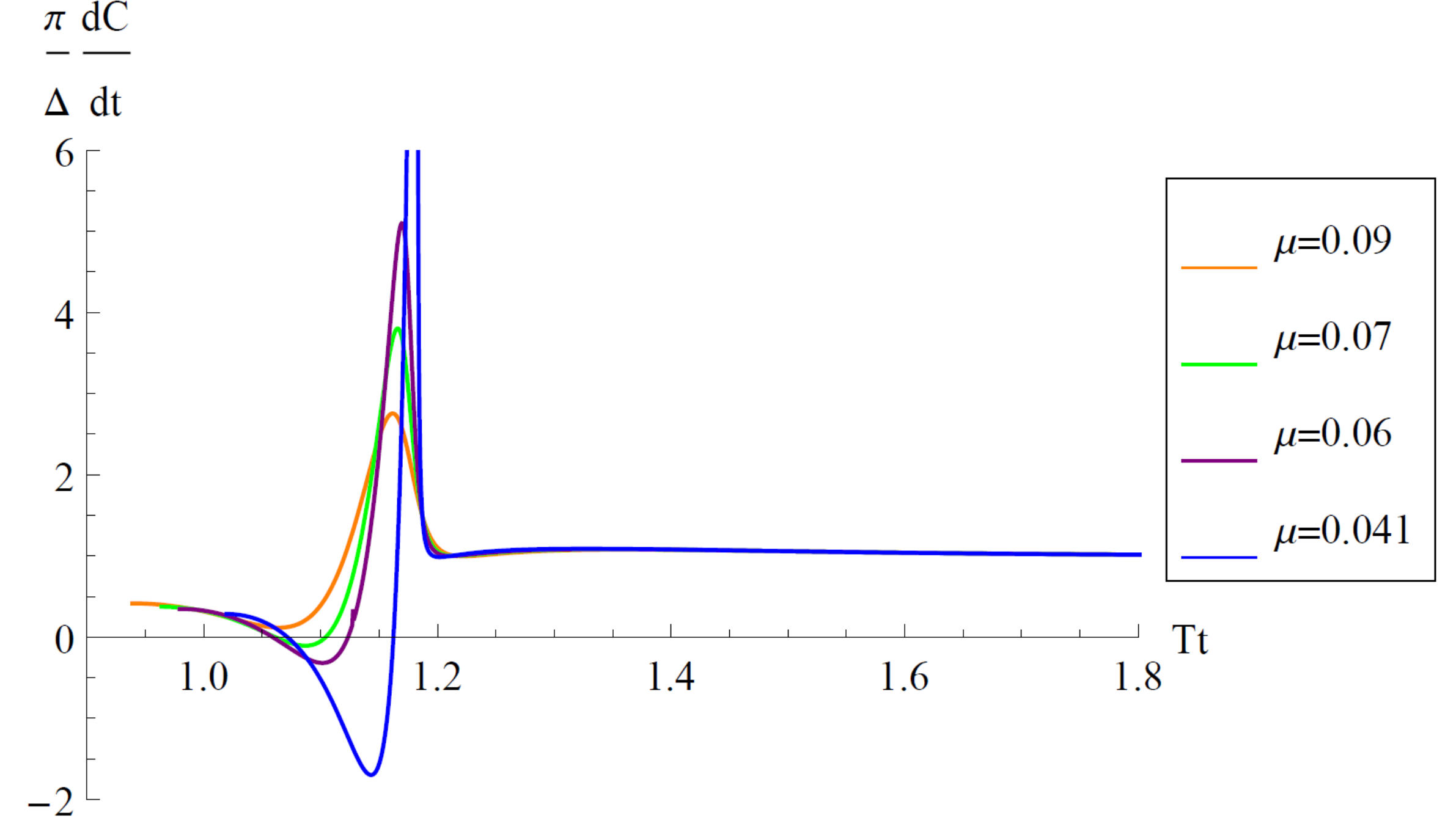}
  	\caption{The time derivative of complexity for the Lovelock black hole (\ref{genericLLBH}) with various couplings $\mu$ in the causal region as a function of the dimensionless time $Tt$. For the left panel $\lambda=-0.001$ and for the right panel $\lambda=-0.2$. We have set $G=\alpha=r_h=\ell=1\,, \omega_5=16\pi$. }
\label{gg} \end{figure}

In Fig.\ref{gg}, we show the numerical results for the time derivative of complexity with various couplings $\mu$. We have fixed $\lambda=-0.001$ in the left panel and $\lambda=-0.2$ in the right panel. Roughly speaking, the time evolution of complexity is very similar to previous cases, except that the coupling constant $\mu$ cannot take values smaller than the critical line $\mu=\lambda^2$. As a matter of fact, as $\mu$ approaches the critical value, the first local maximum becomes higher and sharper. However, it is no longer smoothed out. This is a great difference compared to previous cases.
Nevertheless, there are more similar features. For example, from Fig.\ref{gg}, one observes that the critical time $Tt_c$ is a decreasing function of $\mu$ for a fixed $\lambda$ and the complexity growth rate is a constant at very early times. Indeed, by studying the joint $r_m$ and the complexity growth at very early times, we find that they behave exactly as (\ref{rmLL3}) and (\ref{cpLL3s1}) at leading order.

\section{The time dependence of complexity for charged black holes}\label{chargedsection}
 Now we turn to study the time dependence of complexity for charged Lovelock black holes. The metric function $f(r)$ is determined by (\ref{metricfc}). Depending on the higher order coupling constants, the solution will have either one horizon or two horizons. In this section, we are interested in the latter case. Furthermore, to avoid the existence of an alternative singularity before the inner horizon, the higher order coupling constants should obey certain constraints. We find that for third order Lovelock theories (including Gauss-Bonnet case), the constraints agree precisely with the neutral cases.

\begin{figure}[htbp]
  \centering
  	\subfigure{\includegraphics[width=2.5in]{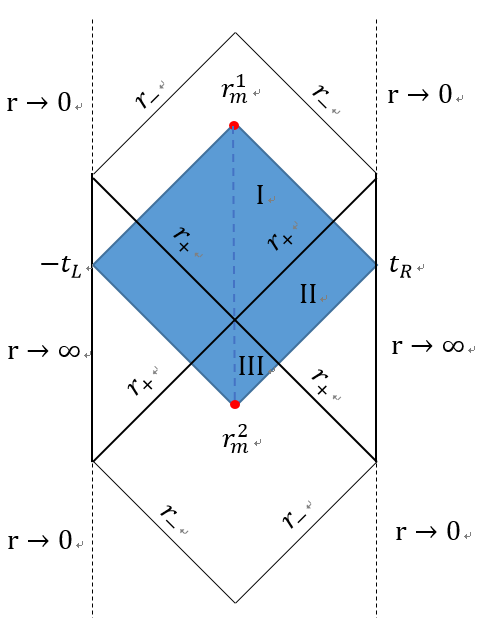}}
  	\caption{The WDW patch for a charged AdS black hole with an inner horizon. There are two joints $r_m^1\,,r_m^2$, evolving with time. }
\label{timecharge}
\end{figure}

 Under the above assumption, the WDW patch for charged Lovelock black holes matches Fig.\ref{timecharge}. Since the WDW patch does not terminate at a singularity, the generalised Gibbons-Hawking-York boundary term evaluated in the neutral case should be absent. Therefore, the relevant gravitational action reduces to the bulk action plus corner terms, namely
\be I_{grav}=I_{bulk}+I_{joint}+\cdots \,,\ee
where again the dotted terms stand for the boundary terms associated to the future and past null boundaries, which however do not have any time dependence. Notice that the time evolution of complexity is implicitly characterized by two joints: the future joint $r_m^1$ and the past joint $r_m^2$. They are determined by
\bea\label{rm12charge}
t\equiv t_R+t_L=2r^*(r_m^1)=-2r^*(r_m^2)\,,
\eea
where $r_-\leq r_m^1, r_m^2\leq r_+$. Evaluating their time derivatives yields
\bea\label{jointevolution}
\frac{dr_m^1}{dt}&=&\frac{1}{2}f(r_m^1)\leq0 \,,\nn\\
\frac{dr_m^2}{dt}&=&-\frac{1}{2}f(r_m^2)\geq0 \,.
\eea
Thus, the past joint $r_m^2$ monotonically increases from the inner horizon $r_-$ to the outer horizon $r_+$ while the future joint $r_m^1$ behaves precisely in the opposite way.

\subsection{Action growth rate}
Despite that the WDW patch for charged black holes looks quite different from the neutral ones, the gravitational action can be computed in a similar way. For example, for bulk action, we split the WDW patch into three pieces and deduce
\bea\label{fullbulkactioncharge}
&&I_{\mathrm{I}}=\fft{\omega_{D-2}}{16\pi G}\int_{r_m^1}^{r_+}dr\,\sqrt{-\bar g}\,\mathcal{L}\,\big[t-2r^*(r) \big]\,,\nn\\
&&I_{\mathrm{II}}=-\fft{\omega_{D-2}}{16\pi G}\int_{r_+}^{\infty}dr\,\sqrt{-\bar g}\,\mathcal{L}\,4r^*(r)\,,\nn\\
&&I_{\mathrm{III}}=\fft{\omega_{D-2}}{16\pi G}\int_{r_m^2}^{r_+}dr\,\sqrt{-\bar g}\,\mathcal{L}\,\big[-t-2r^*(r) \big]\,.
\eea
Then evaluating the time derivative yields
\bea
\frac{dI_{bulk}}{dt}&=&\frac{\omega_{D-2}}{16\pi G}\int^{r_+}_{r_m^1}dr\sqrt{-\bar{g}}\,\mathcal{L}\nn\\
&&-\frac{\omega_{D-2}}{16\pi G}\int^{r_+}_{r_m^2}dr\sqrt{-\bar{g}}\,\mathcal{L}\nn\\
&=&\frac{\omega_{D-2}}{16\pi G}\int^{r_m^2}_{r_m^1}dr\sqrt{-\bar{g}}\,\mathcal{L}\,,
\eea
where we have adopted the relation (\ref{rm12charge}) for the joints.
Using the identity (\ref{bulk}), one finds
\be \frac{dI_{bulk}}{dt}=-\fft{\omega_{D-2}}{16\pi G}\,\sqrt{-\bar g}\mathcal{Q}^{tr}\,\Big|^{r_m^2}_{r_m^1}\equiv -\hat{T}(r)\big[\hat{S}(r)-\hat{S}_A(r)\big]\Big|^{r_m^2}_{r_m^1}\,,
 \ee
where $\hat{S}$ is still the Wald entropy function (\ref{TSfunction}) whilst the function $\hat{S}_A$ is given by
\be\hat{S}_A(r)=\fft{\omega_{D-2}r^{D-2}}{4G}\fft{a(r)a'(r)}{f'(r)} \,.\ee
The existence of this term is associated to the Noether charge of the gauge field
\be \mathcal{Q}_A^{\mu\nu}=-F^{\mu\nu}A_\sigma \xi^\sigma \,. \ee

On the other hand, the relevant joint terms are given by
\bea
I_{joint}&=&\fft{1}{2\pi}\hat{S}(r_m^1)a(r_m^1)+\fft{1}{2\pi}\hat{S}(r_m^2)a(r_m^2) \nn\\
&=&-\fft{1}{2\pi}\hat{S}(r_m^1)\log{\Big(\fft{|f(r_m^1)|}{\alpha^2} \Big)}-\fft{1}{2\pi}\hat{S}(r_m^2)\log{\Big(\fft{|f(r_m^2)|}{\alpha^2} \Big)}\,.
\eea
Taking a derivative with respect to time, one finds
\bea \fft{dI_{joint}}{dt}=\Big[\hat{T}(r)\hat{S}(r)-\fft{\hat{S}'(r)}{4\pi} |f(r)|\log{\Big(\fft{|f(r)|}{\alpha^2} \Big)}\Big]_{r_m^1}^{r_m^2}\,.\eea
Combing all the above results together, we deduce
\bea\label{chargedaction} \fft{dI_{grav}}{dt}&=&\hat{T}(r_m^2)\hat{S}_A(r_m^2)-\hat{T}(r_m^1)\hat{S}_A(r_m^1)\nn\\
&&-\fft{\hat{S}'(r_m^2)}{4\pi} |f(r_m^2)|\log{\Big(\fft{|f(r_m^2)|}{\alpha^2} \Big)}+\fft{\hat{S}'(r_m^1)}{4\pi} |f(r_m^1)|\log{\Big(\fft{|f(r_m^1)|}{\alpha^2} \Big)}\,.\eea

\subsection{Late times and early times}
Before numerically studying the rate of change of complexity, we would like to study it first at late times and early times  half-analytically. This will help us to understand the numerical results better.

\textbf{Late times:} According to (\ref{rm12charge}) or (\ref{jointevolution}), in the late time limit, $r_m^2\rightarrow r_+\,,r_m^1\rightarrow r_-$ to leading order. Consequently, in (\ref{chargedaction}) the first term vanishes because of $a(r_+)=0$ as well as the logarithmic terms. Therefore, to leading order
\be  \fft{dI_{grav}}{dt}\Big|_{late}=-\fft{\omega_{D-2}r_-^{D-2}}{16\pi G}\,a(r_-)a'(r_-)=\Phi_-Q-\Phi_+Q\,.\ee
This reproduces the result first obtained in \cite{Cano:2018aqi} for charged Lovelock black holes. Moreover, it was established \cite{Cano:2018aqi} that the above result can be expressed as
\be\label{chargelate} \fft{dI_{grav}}{dt}\Big|_{late}=2M\, \fft{(1-q^{D-1})(1-q^{D-3})}{1-q^{2(D-2)}}\,,\ee
where $q\equiv r_-/r_+$. Thus, in the uncharged limit $q\rightarrow 0$, one finds $\fft{dI_{grav}}{dt}\rightarrow 2M$ (it should not be confused that when taking the uncharged limit, we always fix the outer event horizon). Interestingly, the result is universal. However, in general it does not match the results of neutral black holes except for the Schwarzschild case. The reason is slightly subtle. In the charged case, the singularity is timelike and the causal structure is very different from the neutral case. This simple fact will not be changed for any given $r_- \neq 0$, even if $r_-$ is sufficiently small. However, for the limiting point $r_-=0$, the causal structure is suddenly changed to the neutral case, which has a spacelike singularity and hence the result should be significantly different. In this sense, there is no reason to believe that in the uncharged limit, the result should reduce to the neutral case, unless a coincidence.

In short, we find a new feature for charged black holes that is not observed in Einstein's gravity:
\be\label{lateinequalityc} \lim_{q\rightarrow 0}\fft{dI_{grav}}{dt}\Big(f(q)\Big)\neq \fft{dI_{grav}}{dt}\Big(\lim_{q\rightarrow 0}f(q)\Big) \,.\ee
 However, we argue that the above results ( normalized by black hole mass ) only differ by a fixed constant at any time $t>t_c$ in the evolution. The reason is with sufficiently small charges, the future joint $r_m^1$ will be exponentially close to the inner horizon $r_-$ during the evolution (see the discussions for early times) so that in the uncharged limit, one expects $r_m^1\rightarrow r_-\rightarrow 0$ (here we mean one first takes the limit $r_m^1\rightarrow r_-$ and then sends $r_-\rightarrow 0$). This implies that
\be\label{unchargedlimit} \lim_{q\rightarrow 0} \fft{dI_{grav}}{dt}=2M-\lim_{q\rightarrow 0} \fft{\hat{S}'(r_m^2)}{4\pi} |f(r_m^2)|\log{\Big(\fft{|f(r_m^2)|}{\alpha^2} \Big)} \,,\ee
where the second term on the r.h.s coincides with that of neutral black holes. We will show that the relation is supported by our numerical results.

To examine the behavior of complexity at late times more carefully, we include the next-to-leading order term. From (\ref{rm12charge}), one finds
\bea\label{rm12}
r_m^1&=&r_-\big(1+c_-e^{-2\pi T_- t}+\cdots\big)\,,\nn\\
r_m^2&=&r_+\big(1-c_+e^{-2\pi T_+ t}+\cdots\big)\,,
\eea
where $c_\pm$ are positive constants given by
\bea\label{cpm}
c_+=\Big(\fft{r_+-r_-}{r_+} \Big)^{\fft{T_+}{T_-}}e^{F(r_+)\int_{r_+}^\infty d\tilde{r}\,H(\tilde{r})}\,,\quad c_-=\Big(\fft{r_+-r_-}{r_-} \Big)^{\fft{T_-}{T_+}}e^{-F(r_-)\int_{r_-}^\infty d\tilde{r}\,H(\tilde{r})}\,,
\eea
where the functions $F(r)\,,H(r)$ are defined by (\ref{chargeFH}). Substituting (\ref{rm12}) into (\ref{chargedaction}), we deduce
\bea
\fft{dI_{grav}}{dt}&=&\Phi_-Q-\Phi_+Q+\sum_\pm \pm 2\pi  c_\pm T_\pm r_\pm \hat{S}'(r_\pm)T_\pm t\, e^{-2\pi T_\pm t}+\cdots \,.
\eea
At the next-to-leading order, the exponential with a smaller exponent will dominate and hence will determine whether the late time limit is approached from above or from below. This depends on the relation between the two temperatures $T_\pm$. However, by computing the temperatures for a variety of dimensions and parameters, we always find $T_+\leq T_-$, where the equality is taken when the solution becomes extremal. This implies that for charged Lovelock black holes the complexity growth rate will generally approach the late time limit from above. Since this is universal to neutral black holes, it might be universal to charged ones as well. Thus, we propose the relation $T_+\leq T_-$ always holds for charged Lovelock black holes. It is of great interest to further investigate its physical consequence and whether this is the case for other charged black holes.\\

\textbf{Early times:} Unlike the neutral case, for charged black holes there is not a critical time below which the rate of change complexity is vanishing. As a matter of fact, for general parameters, the behaviors of the joints $r_m^1\,,r_m^2$ are not universal at very early times. However, for smaller charges, the situation will be different. In this case, the inner horizon will be sufficiently close to the center of spacetimes so that one has according to (\ref{masshorizon})
\be r_-^{D-3}\approx \fft{128\pi^2\ell^2G^2 Q^2}{(D-2)(D-3)\omega^2_{D-2}r_+^{D-1}} \quad\Rightarrow\quad T_-\approx \fft{D-3}{D-1}\fft{T_+}{q^{D-2}} \,.\ee
 Since the temperature on the inner horizon goes as $T_-\sim T_+/q^{D-2}$, the corners will be exponentially close to the inner horizon at very early times, namely
\bea
&&r_m^1=r_-\big(1+c_-\,e^{-2\pi T_- t}+\cdots\big)\,,\nn\\
&&r_m^2=r_-\big(1+c_-\,e^{2\pi T_- t}+\cdots\big)\,,
\eea
where $c_-$ is the positive constant specified by (\ref{cpm}). Here it is worth emphasizing that the result for the future joint $r_m^1$ is valid to any time $t\geq 1/2\pi T_-$ in the evolution because of the negative sign in the exponential.

Clearly, in this case the rate of change in complexity will approximately vanish below a {\it quasi-critical time} $\tilde{t}_c=-\fft{1}{2\pi T_-}\log{c_-}$ at early times. Since $\tilde{t}_c$ should be positive definite, we expect $c_-$ is smaller than unity for smaller charges. This will be ensured in our numerical results. Furthermore, for several solutions in third order Lovelock theories, our numerical results strongly suggest that in the uncharged limit, the {\it quasi-critical time} $\tilde{t}_c$ agrees precisely with the critical time $t_c$ of neutral black holes, though they are defined in quite different ways. While we cannot establish it rigorously, it was shown analytically in \cite{Carmi:2017jqz} for Reissner-Nordstr\o m (RN) black holes.

\subsection{Numerical approach}

Similar to the neutral case, to study the full time dependence of complexity, we need numerically solve the tortoise coordinate at the joints $r_m^1\,,r_m^2$ at first. For this purpose, we follow \cite{Carmi:2017jqz} and introduce two new functions $F(r)\,,H(r)$ as
\bea\label{chargeFH}
&&f(r)= F(r)(r-r_+)(r-r_-)\,,\nn\\
&&H(r)=\fft{F(r_+)r-F(r)r_+}{F(r_+)F(r)r(r-r_+)}-\fft{F(r_-)r-F(r)r_-}{F(r_-)F(r)r(r-r_-)}\,.
\eea
Notice that the two functions are regular on the inner and outer horizons. A useful relation is $F(r_\pm)=4\pi T_\pm/(r_+-r_-)$. The inverse of the metric function $f(r)$ can be nicely written as
\be \fft{1}{f(r)}=\fft{1}{r_+-r_-}\Big[\fft{r_+}{F(r_+)r(r-r_+)}-\fft{r_-}{F(r_-)r(r-r_-)}+H(r) \Big] \,,\ee
so that the tortoise coordinate can be integrated to
\be r^*(r)=\fft{1}{4\pi T_+}\log\Big|\fft{r-r_+}{r}\Big|-\fft{1}{4\pi T_-}\log\Big|\fft{r-r_-}{r}\Big|-\fft{1}{r_+-r_-}\int_r^\infty d\tilde{r}\, H(\tilde{r}) \,.\ee
Again the singular parts on the horizons have been isolated and it is straightforward to solve the tortoise coordinate numerically.

\subsection{Explicit examples}
\begin{figure}[htbp]
  \centering
  \includegraphics[width=220pt]{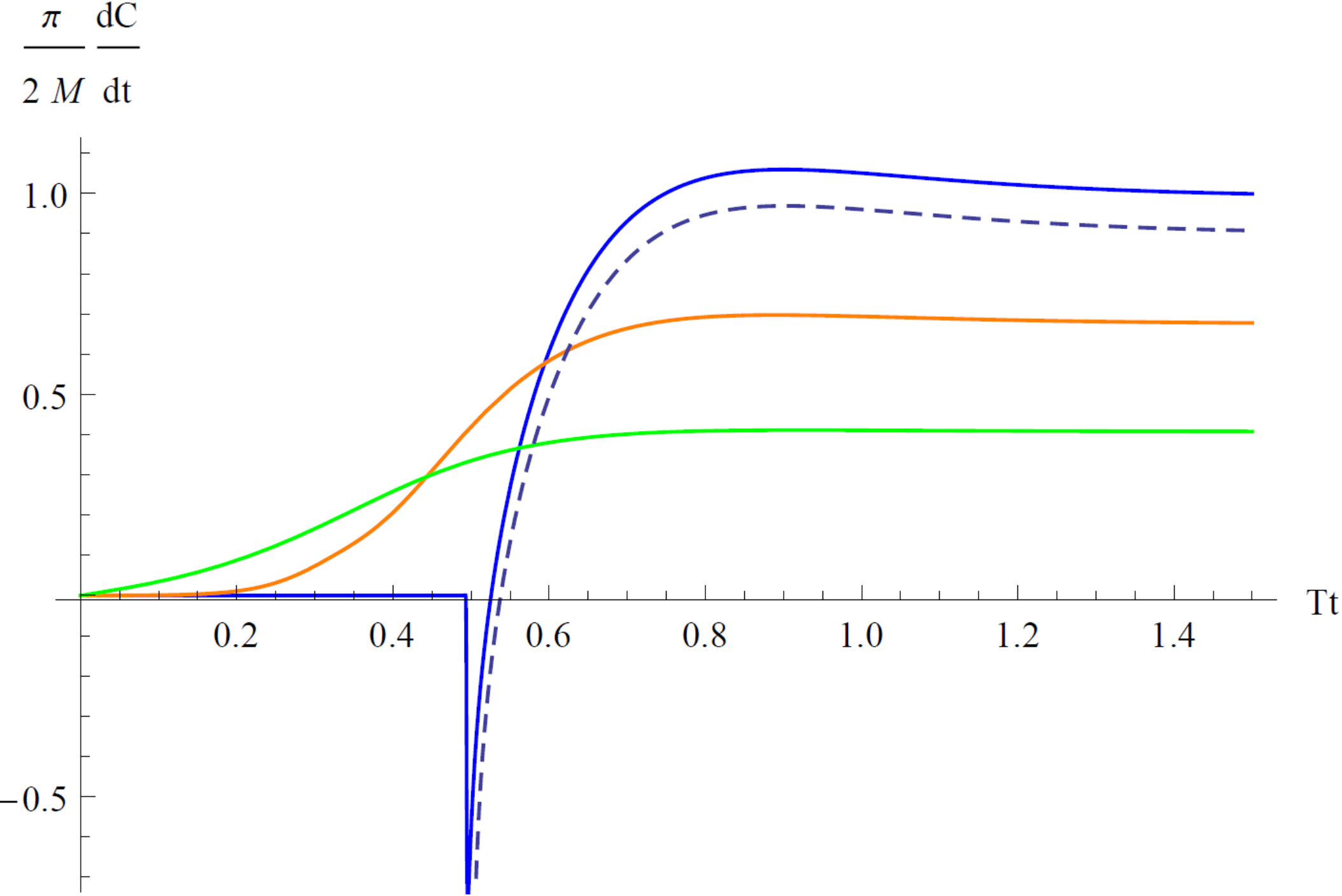}
 \includegraphics[width=210pt]{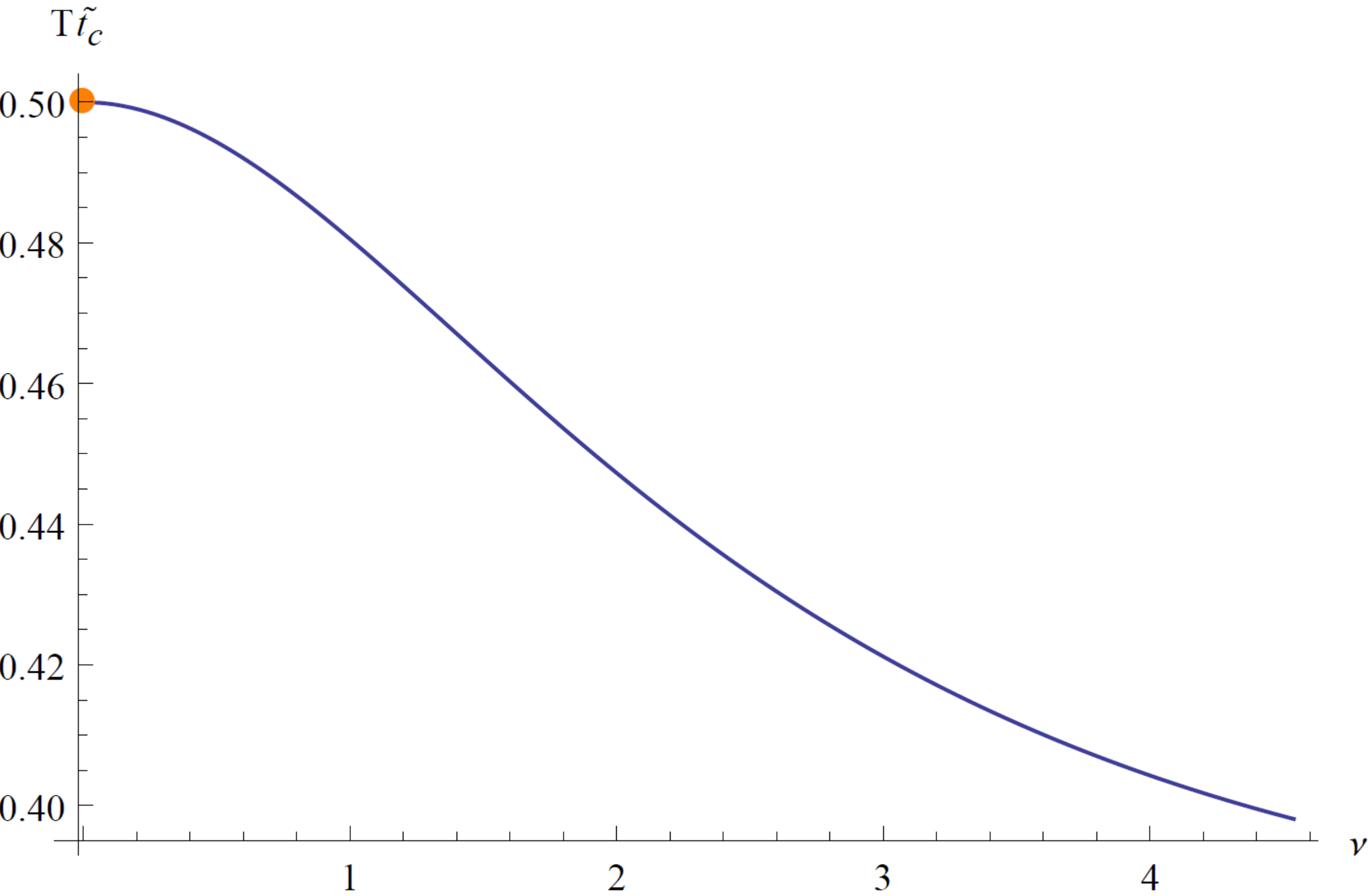}
  	\caption{Left panel: The time derivative of complexity for $D=5$ dimensional RN black holes. $T\ell=1/2$ and $\nu=0.5$ (blue), $\nu=4$ (orange) and $\nu=8$ (green). The dashed line denotes the result of the Schwarzschild black hole with the same size of the $\nu=0.5$ charged black hole. To have a nice presentation, we have slightly moved the result along the vertical axis. Right panel: The quasicritical time $T\tilde{t}_c$ as a function of $\nu$. The orange point denotes the critical time $Tt_c$ of the Schwarzschild black hole. We have set $G=\alpha=\ell=1, \omega_{3}=16\pi$.}
\label{RN} \end{figure}

Using the above results, we are ready to numerically study the full time dependence of complexity for a variety of charged Lovelock black holes. Inspired by the late time result (\ref{chargelate}), we will normalize the complexity growth rate by $2M/\pi$ and plot it as a function of the dimensionless time $Tt$. Here and below, without confusion, we omit the subscript ``+" for the thermodynamical quantities on the outer horizon for convenience. To describe the dependence of complexity growth rate on the electric charges, we introduce a dimensionless quantity $\nu=\Phi/T\ell$, which is related to physical quantities in the boundary CFTs \cite{Carmi:2017jqz}. It follows that $\nu$ is a monotone increasing function of the electric charge $Q$ and $\nu\rightarrow 0$ when $Q\rightarrow 0$. As will be shown later, the complexity growth rate for charged black holes is characterized by $\nu$ and the temperature $T$.\\
\textbf{Reissner-Nordstr\o m black holes}: Our first example is Reissner-Nordstr\o m (RN) black holes, which have
\be f=r^2\ell^{-2}-\fft{16\pi}{(D-2)\omega_{D-2}}\fft{G M}{r^{D-3}}+\fft{128\pi^2}{(D-2)(D-3)\omega_{D-2}^2}\fft{G^2 Q^2}{r^{2(D-3)}} \,.\ee
We focus on the $D=5$ dimensional case. While it has been studied in \cite{Carmi:2017jqz}, it will benefit us to understand the results in higher order theories. In Fig.\ref{RN}, we show the time derivative of complexity for several different values of $\nu$ by fixing the temperature as $T\ell=1/2$. From the figure, we can make a number of observations:\\
$\bullet$ At late times, the complexity growth rate generally approaches the late time limit from above. By numerically computing the ratio $T/T_-$, we find that it is always less than unity by scanning the parameters space. Furthermore, the rate of change of complexity at late times increases as electric charge decreases, consistent with our analytical result (\ref{chargelate}).\\
$\bullet$ For smaller charges, the rate of change of complexity approximately vanishes at very early times. For example, for $\nu=4$, we find $Tt\lesssim 0.2$ whilst $T\tilde{t}_c\simeq 0.4043$. However, the regime becomes larger as the charge becomes smaller. For sufficiently small charge, the regime is enlarged to $t\lesssim \tilde{t}_c$. For instance, for $\nu=0.5$, $t\lesssim \tilde{t}_c\simeq 0.4943T^{-1}$. This has been very close to the critical time of Schwarzschild black holes.\\
$\bullet$ The quasicritical time $\tilde{t}_c$ turns out to be a decreasing function of $\nu$ and it approaches the critical time $t_c$ of Schwarzschild black holes in the uncharged limit, as depicted in the right panel of Fig.\ref{RN}. In fact, it was analytically shown in \cite{Carmi:2017jqz} that in this limit $\tilde{t}_c=t_c=1/2T$.\\
$\bullet$ For smaller charges, the rate of change of complexity develops a local minimum around the quasicritical time. The minimum becomes deeper and sharper as the charge decreases. Thus, we expect that in the uncharged limit, the result will approach to that of Schwarzschild black holes, up to a fixed constant. In fact, for $\nu=0.5$, we have already found that just after the minimum, the result nearly coincides with the Schwarzschild case.

Before proceeding, it is worth emphasizing that the above first two features are valid to charged black holes in higher order theories as well. However, the dependence of the quasicritical time on the charge and the behavior of complexity might be significantly different.\\
\textbf{Charged Gauss-Bonnet black holes: }The second example is charged Gauss-Bonnet black holes, which have
\be f(r)=\fft{r^2}{2\lambda\ell^2}\Big[1-\sqrt{1-4\lambda+\ft{64\pi\lambda\ell^2\, G M}{(D-2)\omega_{D-2}\,r^{D-1}}-\ft{512\pi^2\lambda\ell^2\, G^2 Q^2}{(D-2)(D-3)\omega_{D-2}^2\,r^{2D-4}} }\, \Big]  \,.\ee
\begin{figure}[htbp]
  \centering
  \includegraphics[width=210pt]{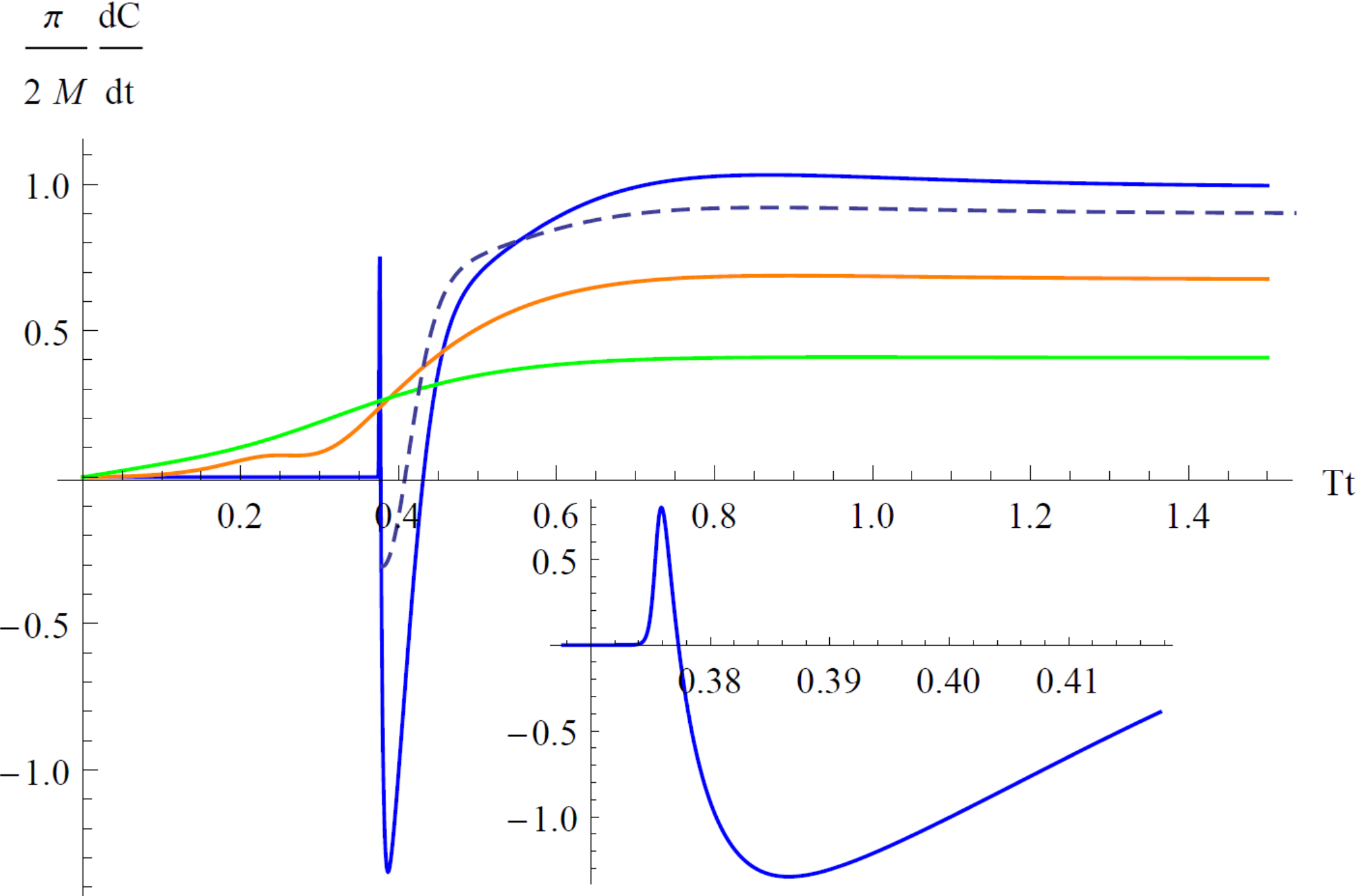}
  \includegraphics[width=210pt]{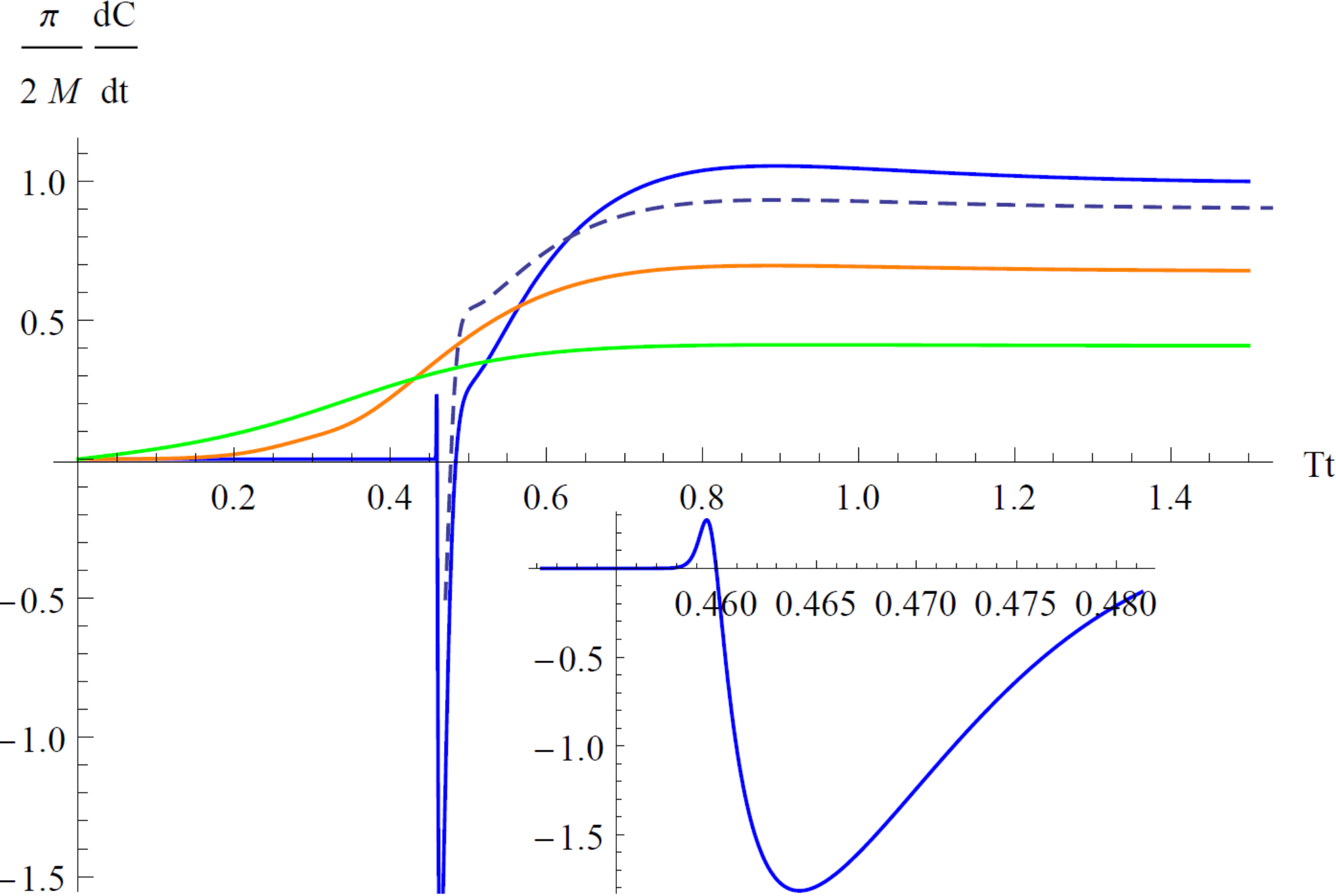}
  	\caption{The time derivative of complexity for charged Gauss-Bonnet black hole in $D=5$ dimension. $\lambda=0.05$ for the left panel and $\lambda=0.01$ for the right panel. In both panels, $T\ell=1/2$ and $\nu=0.5$ (blue), $\nu=4$ (orange), $\nu=8$ (green). The dashed lines denote the results of the neutral black holes with the same size of the $\nu=0.5$ charged black holes. Again, to have a nice presentation, we have slightly moved the neutral results along the vertical axis. We have set $G=\alpha=\ell=1\,,\omega_{3}=16\pi$.}
\label{GB5charge} \end{figure}
In Fig.\ref{GB5charge}, we show the time derivative of complexity for the $D=5$ dimensional solution with different charges. In the left panel $\lambda=0.05$ and in the right panel $\lambda=0.01$. From both panels, we observe that the first three features of RN black holes hold to charged GB black holes as well. However, for smaller charges, the rate of change of complexity first develops a local maximum around the quasicritical time, in addition to the local minimum observed before. As the charge decreases, the local maximum is suppressed and the minimum becomes deeper and sharper. The same phenomenon occurs as well when the Gauss-Bonnet coupling decreases. This is expected since in the RN case, the local maximum does not exist. In the figure, we also show the rate of change of complexity for the neutral black holes, which have the same size of the $\nu=0.5$ charged black holes. It is easy to see that some basic features of the neutral results have already been shown in the $\nu=0.5$ charged results. Moreover, for the latter, the quasicritical time $\tilde{t}_c$ has been very close to the critical time $t_c$ of the neutral black holes. For example, for $\lambda=0.05$, $T\tilde{t}_c\simeq 0.3776\,,Tt_c\simeq 0.3796$ whilst for $\lambda=0.01$, $T\tilde{t}_c\simeq 0.4613\,,Tt_c\simeq 0.4647$. The two only differ by an amount of order $\sim 10^{-3}$. These results suggest that in the uncharged limit, the rate of change of complexity will approach to the neutral results, up to a fixed constant.\\
\textbf{A special charged Lovelock black hole: } Our last example is a $D=7$ dimensional charged black hole in third order Lovelock gravity with coupling constants $\mu=\lambda^2$. The metric function $f(r)$ is given by
\bea\label{specialcharge} f(r)=\fft{ r^2}{\lambda \ell^2}\Big[1-\Big(1-3\lambda+\ft{48\pi\lambda\ell^2 G M}{5\omega_5 r^6}-\ft{96\pi^2\lambda \ell^2 G^2 Q^2}{5\omega_5^2 r^{10}} \Big)^{1/3} \Big]  \,.\eea
\begin{figure}[htbp]
  \centering
  \includegraphics[width=210pt]{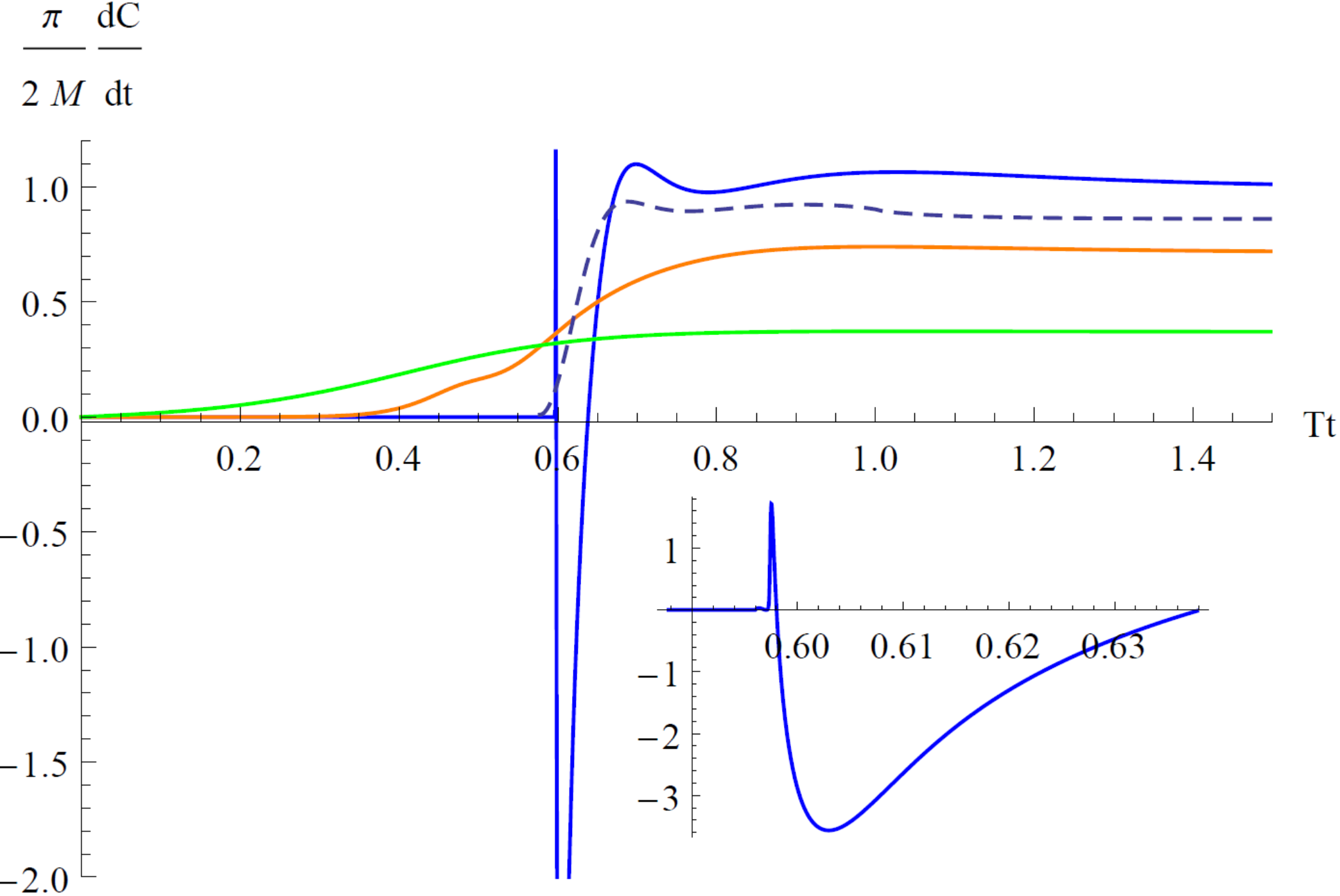}
  \includegraphics[width=210pt]{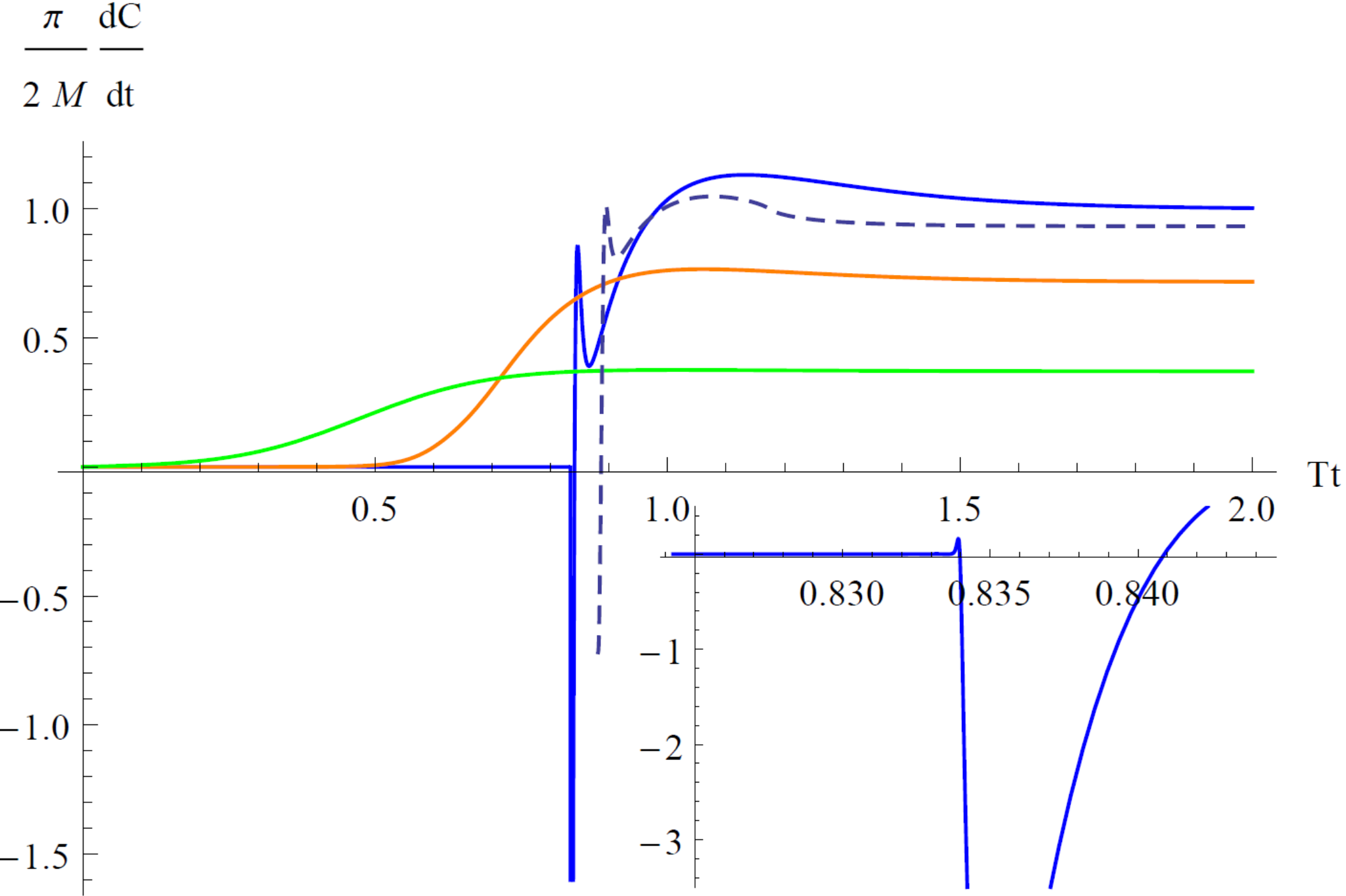}
  	\caption{The time derivative of complexity for the charged black hole (\ref{specialcharge}). $\lambda=0.12$ for the left panel and $\lambda=0.01$ for the right panel. In both panels, $T\ell=1/2$ and $\nu=0.1$ (blue), $\nu=2$ (orange), $\nu=5$ (green). The dashed lines denote the results of the neutral black holes which have the same size of the $\nu=0.1$ charged black holes. To have a nice representation, we have slightly moved the neutral result along the vertical (horizontal) axis in the left (right) panel. We have set $G=\alpha=\ell=1\,,\omega_{5}=16\pi$.}
\label{LL3charge} \end{figure}
In Fig.\ref{LL3charge}, we show the time derivative of complexity for different charges characterized by $\nu=0.1\,,2\,,5$, respectively. In the left panel $\lambda=0.12$ and in the right panel $\lambda=0.01$. From both panels, we again observe the first two features of RN black holes. However, we find that unlike previous cases, the quasicritical time is no longer a decreasing function of $\nu$. Nonetheless, as the GB case, the complexity growth rate first develops a local maximum around the quasicritical time, in addition to the minimum observed before. Furthermore, when the charge or coupling constant $\lambda$ decreases, these extremum behave the same as the GB case. In the figure, we also show the rate of change of complexity for the neutral black holes which have the same size of the $\nu=0.1$ charged black holes. To have a nice representation, we have slightly moved the neutral result along the vertical (horizontal) axis in the left (right) panel. It is immediately seen that the $\nu=0.1$ charged results have already been close to the neutral results. This is shown in the quasicritical time as well. We have for $\lambda=0.12$, $T\tilde{t}_c\simeq 0.5978\,,Tt_c\simeq 0.5772$ whilst for $\lambda=0.01$, $T\tilde{t}_c\simeq 0.8343\,,Tt_c\simeq 0.8312$. In fact, in the uncharged limit, the quasicritical time again approaches the critical time $t_c$ of neutral black holes.

%To end this section, we may conclude that our numerical results support our analytical argument (\ref{unchargedlimit}) about the uncharged limit: the rate of change of complexity (normalized by black hole mass) for charged black holes will reduce to that of neutral black holes, up to a fixed constant.

\section{Conclusion}
In this paper, we study the full time dependence of complexity for Lovelock black holes by using the ``Complexity=Action" (CA) proposal. Though our numerical calculations are performed for the various solutions in third order Lovelock theories, our derivation for the action growth rate in section \ref{sec2} is valid to general higher order gravities. From there, we learn that the action growth rate at late times is essentially determined by the generalised Gibbons-Hawking-York boundary term evaluated at the future singularity. In particular, the ratio of the late time rate to the black hole mass is a pure number, which is independent of the higher order coupling constants and in general is not equal to $2$. As a consequence, the rate of change of complexity generally does not reduce to that of Schwarzschild black holes in the vanishing coupling limit, in spite of that the metric reduces to the latter as well as the gravitational action. The reason is attributed to the inequality (\ref{lateinequality}). However, in the limit the two (the rate normalized by the mass) only differ by a fixed constant at any time $t>t_c$ in the evolution. Here $t_c$ is a critical time, below which the complexity remains a constant.

By including the next-to-leading order term, we find that the complexity growth rate generally approaches the late time limit from above. It implies that any conjectured upper bound on the complexity growth rate given by the late time result will be violated. This extends the result first found for Einstein's gravity \cite{Carmi:2017jqz} to general higher order gravities.

It turns out that at very early times, the behavior of complexity heavily depends on detail of the metric functions. We find that for the various solutions in third order Lovelock theories, it is either a constant or logarithmically diverges. However, a generic feature is the critical time $t_c$ turns out to be a decreasing function of the higher order couplings, implying that the complexity evolves faster than that of Schwarzschild black holes.

%In addition, during the evolution there exists a characteristic feature for the various solutions in third order theories, except the reduced cases, i.e. the Gauss-Bonnet and Einstein's gravity: after the early times the rate of change of complexity first develops a new local maximum, in addition to the old one close to the late times. In general, this local maximum becomes higher and sharper for smaller couplings and will disappear in the vanishing coupling limit. Our numerical results suggest that the complexity growth will reduce to the Schwarzschild case, up to a fixed constant.

We further study the time dependence of complexity for charged Lovelock black holes with an inner horizon. With sufficient charge, the complexity roughly behaves the same as that in Einstein's gravity. However, for smaller charges, the two have some significant differences. In particular, in the uncharged limit, the rate of change of complexity at late times is universal
\be \lim_{q\rightarrow 0}\fft{d\mathcal{C}}{dt}=\fft{2M}{\pi} \,,\ee
in contrast to the non-universal results for the neutral black holes. The reason for the mismatch can be attributed to the inequality (\ref{lateinequalityc}), which results from a sudden change of the causal structure. In fact, we find that the two differ by a fixed constant in the whole time evolution.

%Furthermore, by scanning the parameters space, we find that the complexity growth rate always approaches the late time limit from above. This motivates us to propose a universal inequality for the temperatures $T_\pm$ associated to the outer (``+") and inner (``-") horizons: $T_+\leq T_-$, where the equality is taken when the solution becomes extremal. It will be interesting to further explore its physical consequence and whether this is the case for other charged black holes.

\section*{Acknowledgments}
We are grateful to Jingyi Zhang for useful discussions. Z.Y.~Fan is supported in part by the National Natural Science Foundations of China (NNSFC) with Grant No. 11805041, No. 11873025 and No. 11575270. H.Z.~Liang is supported in part by NNSFC No. 11873025.\\

\appendix
\section{The Noether charge for third order Lovelock gravities}
Our discussions in section (\ref{bulkrate}) on the action growth rate for general higher order gravities strongly relies on the identity (\ref{bulk}), which connects the bulk Lagrangian and the Noether charge associated to a stationary black hole. In fact, from the Wald-Iyer formalism \cite{wald1,wald2}, the Noether charge for a higher order gravity of the form
$\mathcal{L}=\mathcal{L}(g_{\mu\nu};R_{\mu\nu\rho\sigma})$ can be derived as
\be\label{Noethertotapp} \mathcal{Q}^{\mu\nu}=-2E^{\mu\nu\rho\sigma}\nabla_\rho \xi_\sigma-4\xi_\rho \nabla_\sigma E^{\mu\nu\rho\sigma}  \,,\ee
where $\xi$ is a Killing vector and
\be E^{\mu\nu\rho\sigma}\equiv \fft{\partial \mathcal{L}}{\partial R_{\mu\nu\rho\sigma}} \,.\ee
Since this quantity is of great importance in our practical calculations in section \ref{EGBcomp}, \ref{LL3comp} and \ref{chargedsection}, here we shall discuss it further, in particular for the third order Lovelock theories.

Certainly, it is straightforward to derive the tensor $ E^{\mu\nu\rho\sigma}$ for a given gravitational Lagrangian but one should remember that by definition, the tensor $E_{\mu\nu\rho\sigma}$ shares indices symmetries of the Riemainnian tensor, namely
\be E_{\mu\nu\rho\sigma}=E_{[\mu\nu][\rho\sigma]}=E_{\rho\sigma\mu\nu}\,,\quad E_{[\mu\nu\rho]\sigma}=0 \,.\ee
This is important to avoid any mistake in the derivations (it can also be used to check whether the result is trustworthy or not). As simple examples, we would like to provide the results for some low-lying curvature polynomials: $\mathcal{L}=R\,,R_{\mu\nu}^2\,,R_{\mu\nu\rho\sigma}^2$. We have
\bea
&&R\,:\qquad E^{\mu\nu\rho\sigma}=g^{\rho[\mu}g^{\nu]\sigma}=\ft12\big(g^{\mu\rho}g^{\nu\sigma}-g^{\mu\sigma}g^{\nu\rho} \big)\,,\nn\\
&&R_{\mu\nu}^2\,:\qquad E^{\mu\nu\rho\sigma}=g^{\rho[\mu}R^{\nu]\sigma}-g^{\sigma[\mu}R^{\nu]\rho}\,,\nn\\
&&R_{\mu\nu\rho\sigma}^2\,:\qquad E^{\mu\nu\rho\sigma}=2R^{\mu\nu\rho\sigma}\,.
\eea
Here and below, it should not be confused that we always use $E^{\mu\nu\rho\sigma}$ to denote the tensor associated to a single curvature polynomial. From these results, it is straightforward to derive the Noether charge (\ref{EGBNoether}) for Einstein-Gauss-Bonnet gravity. We rewrite the result as follows
 \bea
\mathcal{Q}^{\mu\nu}_{EGB} = -2\Big(\nabla^{[\mu}\xi^{\nu]}
+\ft{2\lambda\,\ell^{2}}{(D-3)(D-4)}\,\big(R\,\nabla^{[\mu}\xi^{\nu]}-4 R^{\sigma[\mu}\nabla_\sigma \xi^{\nu]}+ R^{\mu\nu\sigma\rho}\nabla_\sigma \xi_\rho \big)\Big)\,.
\eea
For the third order Lovelock gravities (\ref{LL3lagrangiantot}), the Noether charge can be formally written as
\be \mathcal{Q}^{\mu\nu}=\mathcal{Q}^{\mu\nu}_{EGB}+\ft{\mu\,\ell^4}{3(D-3)(D-4)(D-5)(D-6)}\,\mathcal{Q}^{\mu\nu}_{3} \,,\ee
where $\mathcal{Q}^{\mu\nu}_3$ stands for the Noether charge for the cubic Lagrangian $\mathcal{L}_3$ in (\ref{LL3lagrangian}). To calculate $\mathcal{Q}^{\mu\nu}_3$, one needs first calculate the corresponding tensor $E^{\mu\nu\rho\sigma}$ for each curvature polynomial of $\mathcal{L}_3$. Since the derivations are simple and straightforward, we just present the final results in the following
\bea\label{NoetherLL3}
&& R^3\,:\qquad E^{\mu\nu\rho\sigma}=3R^2 g^{\rho[\mu}g^{\nu]\sigma} \,,\nn\\
&& RR_{\mu\nu\rho\sigma}^2\,: \qquad E^{\mu\nu\rho\sigma}=2R R^{\mu\nu\rho\sigma}+g^{\rho[\mu}g^{\nu]\sigma} R^2_{\alpha\beta\gamma\delta}\,,\nn\\
&& RR_{\mu\nu}^2\,: \qquad E^{\mu\nu\rho\sigma}=g^{\rho[\mu}g^{\nu]\sigma} R^2_{\alpha\beta}+R\big(g^{\rho[\mu}R^{\nu]\sigma} -g^{\sigma[\mu}R^{\nu]\rho}\big)\,,\nn\\
&& R^{\mu\nu\rho\sigma}R_{\mu\rho}R_{\nu\sigma}\,: \qquad E^{\mu\nu\rho\sigma}=
R^{\rho[\mu}R^{\nu]\sigma}+R_{\alpha\beta}\big(g^{\rho[\mu}R^{\nu]\alpha\sigma\beta} -g^{\sigma[\mu}R^{\nu]\alpha\rho\beta}\big)\,,\nn\\
&& R^{\mu\nu}R_{\mu\sigma}R_{\nu}^{\sigma}\,: \qquad E^{\mu\nu\rho\sigma}=\fft32\big(
g^{\rho[\mu}R^{\nu]}_\alpha R^{\alpha\sigma} -g^{\sigma[\mu}R^{\nu]}_\alpha R^{\alpha\rho}\big)\,,\nn\\
&& R^{\mu\nu\rho\sigma}R_{\rho\sigma\nu\lambda}R_{\mu}^{\lambda}\,: \qquad E^{\mu\nu\rho\sigma}=2R^{[\mu}_{\alpha}R^{\nu]\alpha\rho\sigma}-\fft12\big(
g^{\rho[\mu}R^{\nu]\alpha\beta\gamma} R^{\sigma}_{\,\,\,\,\alpha\beta\gamma} -g^{\sigma[\mu}R^{\nu]\alpha\beta\gamma} R^{\rho}_{\,\,\,\,\alpha\beta\gamma}\big)\,,\nn\\
&& R^{\mu\nu}_{\quad \alpha\rho}R^{\alpha\beta}_{\quad \nu\sigma }R^{\rho\sigma}_{\quad \mu\beta}\,: \qquad E^{\mu\nu\rho\sigma}=\fft32\big(R^{\alpha\sigma\beta[\mu}R^{\nu]\,\,\,\,\rho}_{\,\,\,\,\alpha\,\,\,\,\beta}
-R^{\alpha\rho\beta[\mu}R^{\nu]\,\,\,\,\sigma}_{\,\,\,\,\alpha\,\,\,\,\beta}  \big)\,,\nn\\
&& R^{\mu\nu\alpha\beta}R_{\alpha\beta\rho\sigma}R^{\rho\sigma}_{\quad \mu\nu}\,: \qquad E^{\mu\nu\rho\sigma}=3 R^{\mu\nu\alpha\beta}R_{\alpha\beta}^{\quad \rho\sigma}\,.
\eea
With these results in hand, it is easy to compute $\mathcal{Q}^{\mu\nu}_3$ via (\ref{Noethertotapp}).

\end{document}